\documentclass[lettersize,journal]{IEEEtran}
\usepackage{algorithmic}
\usepackage{algorithm}
\usepackage{array}
\usepackage[caption=false,font=normalsize,labelfont=sf,textfont=sf]{subfig}
\usepackage{textcomp}
\usepackage{stfloats}
\usepackage{url}
\usepackage{verbatim}
\usepackage{graphicx}
\usepackage{cite}
\usepackage{amsmath,amssymb,amsfonts,amsthm}
\usepackage{bm}
\usepackage{mathtools}
\usepackage{booktabs}
\usepackage{tabularx}
\usepackage{xcolor}
\usepackage{hyperref}
\usepackage{cleveref}
\usepackage{mathrsfs}

\begin{document}
	
	\title{On Privacy-Preserving Image Transmission in Low-Altitude Networks: A Swin Transformer-Based Framework with Federated Learning}
	
	\author{Kexin~Zhang,
		Lixin~Li,
		Yuna~Yan,
		Xin~Zhang,
		Wensheng~Lin,
		Rui~Li,
		Dongwei~Zhao,
		and~Zhu~Han
		\thanks{K. Zhang, L. Li (corresponding author, e-mail: lilixin@nwpu.edu.cn), Y. Yan, X. Zhang, and W. Lin are with the School of Electronics and Information, Northwestern Polytechnical University, Xi'an 710129, China. L. Li and W. Lin are also with DecoreX Intelligent Technologies Co., Ltd., Xi'an 710075, China.}
		\thanks{R. Li is with Samsung AI Center, Cambridge CB1 2JH, UK.}
		\thanks{D. Zhao is with the 208th Research Institute of China North Industries Group Corporation, Beijing 102202, China.}
		\thanks{Z. Han is with the Department of Electrical and Computer Engineering, University of Houston, Houston, TX 77004, USA.}}
	
	\maketitle
	
	\begin{abstract}
		The rapid development of low-altitude economy has driven the proliferation of Unmanned Aerial Vehicle (UAV) applications, including logistics, inspection, and emergency response. However, transmitting high-volume image data from UAVs to ground stations faces significant challenges due to limited bandwidth and stringent privacy requirements. To address these issues, a Semantic Communication (SC) framework based on Federated Learning (FL) is proposed for efficient and privacy-preserving image transmission. A Swin Transformer-based Semantic Communication (STSC) architecture is designed to extract multi-scale semantic features under constrained bandwidth conditions. Dedicated communication and computing nodes are deployed on UAVs to enhance real-time coverage and flexibility. Meanwhile, a FL mechanism enables global model training across distributed devices without sharing raw data, thus preserving user privacy. Simulation experiments conducted on the CIFAR-10 dataset demonstrate that the proposed STSC framework achieves at least 5.7 dB improvement in Peak Signal-to-Noise Ratio (PSNR) compared to DeepJSCC baselines, while also showing superior convergence and generalization performance. The framework effectively integrates UAV-assisted deployment with SC and privacy protection, offering a practical solution for bandwidth-constrained image transmission in low-altitude networks.
	\end{abstract}
	

\section{Introduction}
\label{sec:introduction}

The rapid advancement of low-altitude economy has accelerated the
deployment of Unmanned Aerial Vehicles (UAVs) across diverse domains,
including logistics delivery, infrastructure inspection, agricultural
monitoring, and emergency response. As essential nodes in low-altitude
networks, UAVs continuously capture and transmit high-volume image data
to ground stations or command centers for real-time analysis and
decision-making. 
However, challenges such as limited bandwidth, unreliable 
air-to-ground links, and constrained energy resources 
complicate the efficient transmission of images. In 
particular, RF transmission constitutes the dominant 
source of energy consumption on battery-powered UAV 
platforms, making bandwidth-efficient 
transmission a critical factor for extending operational 
endurance.

Traditional communication architectures separate source and channel
coding, where images are first compressed and then protected against
errors. Although optimal in theory for infinite-length messages, this
approach introduces significant delay and complexity, which conflicts
with the low-latency demands of low-altitude applications.
Additionally, it treats all bits equally, neglecting the prioritization
of semantically substantial content, and places a heavy computational
load on UAVs, which have limited resources.

To address these issues, Semantic Communication (SC) focuses on
transmitting meaningful information rather than exact bits~\cite{ref1,ref10}. In
UAV-enabled low-altitude networks, SC allows UAVs to extract key
semantic features from images and transmit only relevant content to
ground stations, significantly reducing bandwidth consumption and delay.
Building on this, we propose a Swin Transformer-based Semantic
Communication (STSC) system for efficient image transmission. The Swin
Transformer's hierarchical structure enables multi-scale semantic
feature extraction with low computational overhead, making it suitable
for resource-constrained UAV deployment. Additionally, in low-altitude
networks, multiple UAVs and ground units collaborate to process data,
raising concerns about data privacy and security. For instance, UAVs
performing inspection tasks may capture sensitive industrial facilities,
while those in logistics may record private residential areas. To
safeguard sensitive information, we integrate Federated Learning (FL)
into the semantic communication framework. Instead of transmitting raw
images, UAVs and user devices train local models and share only model
updates, preserving data privacy while enabling collaborative learning.
In our preliminary work~\cite{ref0}, we explored the potential of Swin Transformer architectures for federated semantic image communication. Building upon that foundation, the present study further develops a complete system design tailored to low-altitude UAV networks, encompassing multi-condition channel adaptation, heterogeneous data robustness, and rigorous privacy analysis.

Despite these advances, directly applying existing FL-based semantic communication methods 
to low-altitude UAV image transmission remains non-trivial, for three concrete reasons.
First, prior works in this space~\cite{ref24,ref25} primarily target text and audio modalities, 
whose compact token representations and relatively homogeneous data statistics are structurally 
distinct from the high-dimensional, spatially structured distributions of UAV imagery.
Second, for image-oriented baselines such as DeepJSCC~\cite{ref2}, the convolutional backbone 
lacks hierarchical multi-scale feature extraction, which fundamentally limits its capacity to 
benefit from federated aggregation: as demonstrated in Section~\ref{sec_5}, the 
federated DeepJSCC global model yields only marginal reconstruction improvement over locally 
trained models and requires approximately 2000 training epochs to converge, 
compared to 60 epochs for the proposed STSC framework.
Third, UAV air-to-ground channels span a wide range of propagation conditions, from strong line-of-sight during open-air cruising to severe multipath fading in complex terrain~\cite{zeng2016energy}; moreover, emerging 6G scenarios may further introduce heavy-tailed impulsive interference that departs from conventional Gaussian assumptions~\cite{LiChannel2026}, a diversity that existing FL-semantic-communication designs developed for terrestrial scenarios do not account for.
These gaps motivate the dedicated design of the STSC framework, which jointly 
addresses multi-scale semantic representation for image modalities and 
federated privacy-preserving training.

The proposed system includes communication and computation modules on
UAVs for real-time semantic processing at the edge. UAVs extract local
semantic knowledge from images and participate in federated training to
build a global model for semantic encoding. This FL-enabled semantic
communication framework is suitable for bandwidth-limited,
energy-constrained, and privacy-sensitive low-altitude network
environments.

In summary, the main contributions of this work are:
\begin{itemize}
	
	\item We investigate the integration of Swin Transformer as 
	a joint source-channel encoder-decoder for UAV image 
	transmission in low-altitude networks. The hierarchical 
	patch-merging pipeline and shifted window attention mechanism 
	are configured to meet the compression and latency 
	requirements of bandwidth-constrained UAV uplinks, and the 
	encoder-decoder pair is optimized end-to-end under channel 
	noise, distinguishing the system from Swin Transformer 
	applications in noise-free visual recognition tasks.
	
	\item We develop a cross-silo federated learning framework 
	that integrates STSC across distributed UAV edge devices, 
	enabling collaborative model training without exposing raw 
	imagery. The framework is systematically evaluated under 
	both IID and non-IID data distributions to characterize 
	the impact of data heterogeneity arising from geographically 
	diverse UAV missions on federated convergence and 
	reconstruction quality.
	
	\item Experiments under AWGN, Rician, and Rayleigh fading 
	channels show that the proposed framework achieves at least 
	5.7~dB PSNR improvement over DeepJSCC baselines, and the 
	federated global model attains 2--3~dB gain over locally 
	trained models with significantly faster convergence, 
	providing empirical evidence that Swin Transformer-based 
	semantic coding is better suited to federated image 
	transmission than convolutional alternatives.
	
\end{itemize}

The remainder of this paper is organized as follows. Section \ref{sec_2} reviews related work. Section \ref{sec_3} introduces the proposed federated image semantic communication framework and problem formulation. Section \ref{sec_4} details the Swin Transformer-based federated learning strategy. Section \ref{sec_5} presents simulation results and performance evaluation. Finally, Section \ref{sec_6} concludes the paper.

\section{Related Works}
\label{sec_2}
\subsection{UAV-Assisted Low-Altitude Communications}
The rapid development of low-altitude economy has driven extensive research on UAV-assisted communications. The authors of \cite{UAV3} explored UAV systems with reconfigurable intelligent surfaces for air-to-ground communications, introducing fading-shadowing models to address related challenges. 
In addition, Lin et al. \cite{LinUAV2022} investigated the timeliness optimization of UAV lossy communications for IoT applications, highlighting the critical trade-off between data freshness and communication reliability under constrained UAV resources.
Moreover, resource-efficient UAV deployment strategies, including optimal trajectory and power control for multi-type UAV aerial base stations \cite{LiUAV2021}, further underscore the importance of communication-aware design in low-altitude networks.
However, UAVs have limited power and computing capabilities, making traditional coding pipelines impractical. Efficient image transmission is critical yet challenging, as UAVs must allocate limited energy not only to communication but also to flight and auxiliary functions \cite{UAV2}.
Recent advances in aerial edge computing, such as distributed user pairing and computation offloading in aerial edge networks \cite{LiangUAV2024}, have demonstrated the feasibility of deploying computational resources on UAV platforms, motivating our edge-based semantic processing architecture.
Low-altitude networks often rely on UDP-based protocols for real-time data transmission in time-sensitive applications such as inspection, logistics, and emergency response \cite{UAV4}. However, the lack of ARQ mechanisms in UDP results in poor performance of traditional source-channel coding under unreliable channel conditions \cite{UAV5}. While separation-based coding is optimal for infinite-length messages, it introduces delays and complexity that fail to meet the low-latency demands of low-altitude applications. These limitations underscore the need for robust, low-latency transmission frameworks designed explicitly for UAV-assisted low-altitude networks.
Beyond transmission efficiency, the convergence of sensing and communication in low-altitude networks has also attracted increasing attention; for instance, Zheng \emph{et al.}~\cite{ZhengLAN2026} proposed an optimal transport framework for integrated sensing and communication with joint resource allocation for cooperative communication and non-cooperative localization, further highlighting the multi-dimensional design challenges in such networks.

\subsection{Image Semantic Communication}
Traditional image transmission systems rely on separate source and channel coding. Source coding methods, such as JPEG and JPEG 2000, compress images at the pixel level, while channel coding techniques, such as LDPC and turbo codes, provide error protection. However, this separation approach has notable drawbacks: (1) it ignores the semantic relevance important for human perception, (2) it suffers from limited adaptability and high computational complexity, and (3) it performs poorly under varying channel conditions.

Deep learning-based semantic communication addresses these issues by jointly optimizing source and channel coding~\cite{ref11,ref12}. Instead of focusing on pixel-level accuracy, these systems encode images into compact semantic representations and reconstruct them at the semantic level, improving efficiency and robustness.

Recent developments include: deep JSCC schemes that avoid the "cliff effect" and degrade gracefully under poor conditions \cite{ref2}; adaptive rate control based on channel state and content \cite{ref16}; multi-user and multimodal extensions for tasks like image retrieval and visual question answering \cite{ref17}; channel feedback mechanisms for enhanced performance \cite{ref18}; and novel architectures like the WITT Transformer \cite{ref19}. Practical validation on FPGA platforms \cite{ref20} has demonstrated superior performance compared to traditional methods, such as 256-QAM, particularly in low Signal-to-Noise Ratio (SNR) scenarios. Moreover, Zhang \emph{et al.}~\cite{Zhang} proposed a generative AI-aided semantic successive refinement framework, and Yan \emph{et al.}~\cite{Yan} investigated adaptive semantic generation with NOMA-based interference management for 6G networks.

\subsection{Multi-user Semantic Communications on FL}
The above methods for image transmission face a critical privacy challenge: they require centralized training on raw datasets containing sensitive personal information (e.g., facial images), creating significant privacy risks when data is uploaded to cloud servers.

FL offers a privacy-preserving alternative by enabling distributed model training without accessing users' raw data \cite{ref22}. This approach allows semantic communication systems to benefit from collaborative learning while maintaining local data privacy.
Moreover, recent advances in federated edge learning have addressed the challenges of heterogeneous data and resource allocation, such as joint client scheduling and wireless resource allocation under non-IID data distributions \cite{YinFL2024}, which motivates our investigation of federated semantic communication under data heterogeneity.
Similarly, federated learning has been applied to energy-efficient configuration optimization for multiple intelligent reflecting surfaces \cite{LiFL2022}, demonstrating the versatility of FL-based approaches in communication system design.

Recent FL-based semantic communication research has primarily focused on non-image modalities. Shi \emph{et al.} \cite{ref23} proposed joint edge intelligence architectures for offloading semantic processing tasks. Tu \emph{et al.} \cite{ref24} implemented Transformer-based systems for text communication, while Tong \emph{et al.} \cite{ref25} developed wav2vec-based autoencoders for audio transmission. However, FL-based semantic communication for image transmission remains underexplored.

These studies primarily focus on FL-based semantic communication for voice and text, while overlooking its application in image transmission. This work bridges that gap by integrating image-based semantic communication with federated learning to enhance semantic aggregation, improve image reconstruction, and preserve user privacy.

\begin{figure*}[!t]
	\centering
	\includegraphics[width=6.1in]{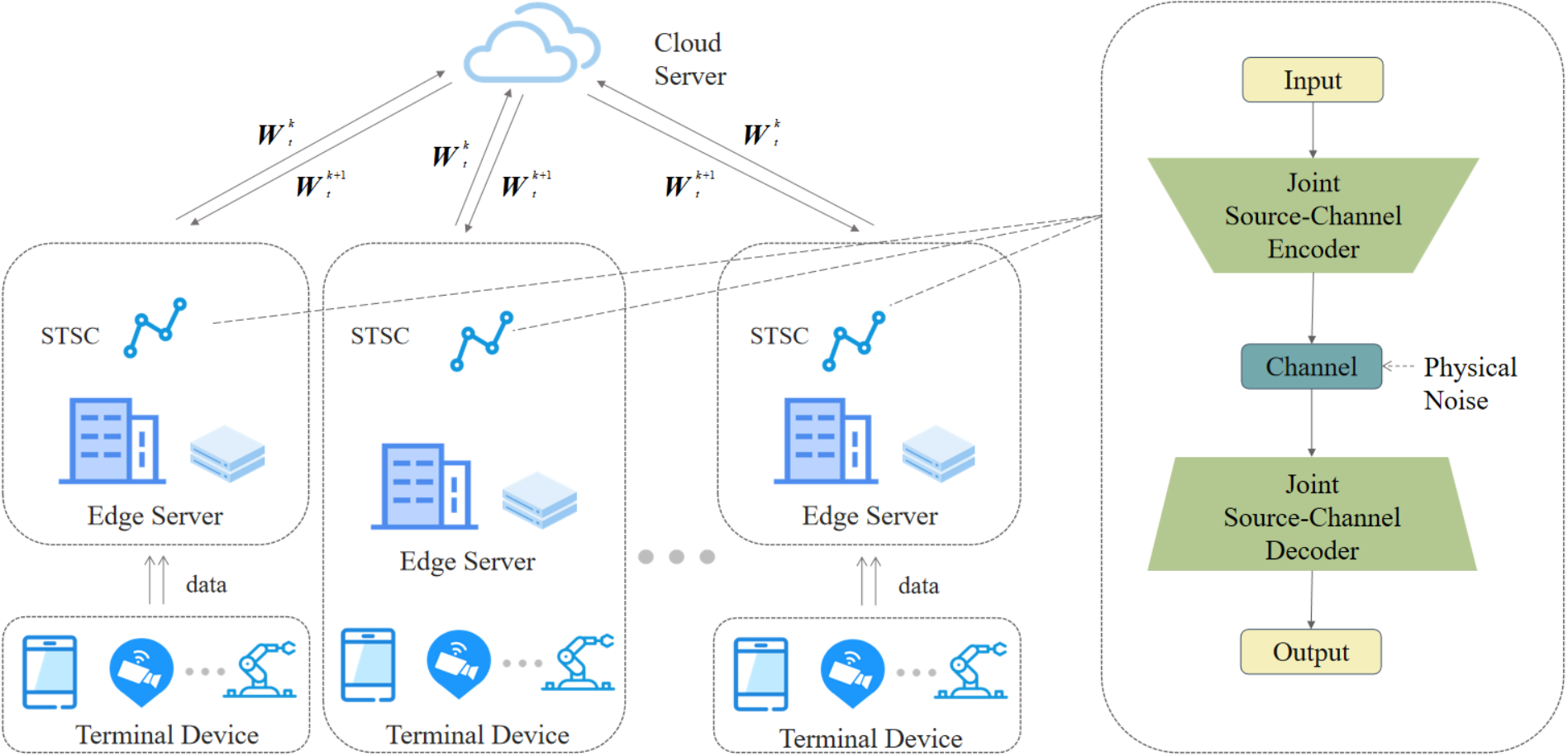}
	\caption{The overall architecture of the multi-user federated learning semantic communication system.}
	\label{fig_1}
\end{figure*}

\subsection{Swin Transformer} 
As the Transformer architecture \cite{ref26} has achieved remarkable success in language processing, researchers have also explored its potential for visual tasks. Vision Transformers (ViT) \cite{ref27} successfully adapt the Transformer architecture to image classification tasks. However, ViT's focus on global information processing limits its ability to capture local features, resulting in quadratic computational complexity with respect to image size. Swin Transformer \cite{ref28} addresses these issues through hierarchical processing with shifted windowing, reducing computational complexity while enabling both local and cross-window attention. Given these advantages, we adopt Swin Transformer as the backbone architecture for our semantic communication network.

\section{System Setup}
\label{sec_3}
A multi-user federated learning semantic communication system is considered in this paper, as illustrated in Fig.\ref{fig_1}, comprising a cloud server and multiple clients. Each client can be either an edge server or a terminal device. If the client is an edge server, it can perform computational tasks; therefore, STSC training and inference are carried out locally. If the client is a lightweight terminal device with limited or no computational power, it is supported by a corresponding trusted edge server, which handles data and training on its behalf. The federated training process is detailed as follows:

\begin{enumerate}
	\item The cloud server initializes and distributes the STSC model to all clients. Each client performs n epochs of local training and then transmits the updated parameters to the server for Federated Averaging aggregation \cite{ref29}. This process repeats until convergence;
	\item During deployment, clients process semantic representations through the trained STSC model for joint source-channel encoding and transmit the encoded output via uplink channels;
	\item The transmitted signal is corrupted by channel noise. Upon reception, the decoder reconstructs the original message using the STSC decoding module. The receiver can be either a cloud server or another client.
\end{enumerate}

The semantic communication system has three main components: a joint source-channel encoder, a physical channel, and a joint source-channel decoder. Each of these three components will be described in the following subsections, respectively.


\subsection{Joint Source-Channel Encoder} 
As shown in Fig.\ref{fig_1}, when the transmitter sends an image, the joint source-channel encoder performs feature extraction. The aim is to focus on informative semantic information and remove irrelevant semantic information in non-target areas. After encoding by the joint source-channel encoder, the semantic features of the image are obtained and converted into a complex symbols (vector) suitable for transmission on the physical channel, i.e.,

\begin{equation}
	\label{Eq_1}
	\boldsymbol{s}=T\left(\boldsymbol{x}, \boldsymbol{\varphi}_\alpha\right),
\end{equation}
where  $ \boldsymbol{x} $ is an $M \times N$  dimensional vector,  $T(\cdot)$ represents joint source-channel encoding network, and $ \boldsymbol{\varphi}_\alpha $  is the set of parameters for the corresponding encoding network.

\subsection{Physical Channel}
\label{sec:channel}
The physical channel covers typical UAV air-to-ground propagation conditions. The received signal is modeled as
\begin{equation}
	\label{Eq_channel}
	\boldsymbol{y} = h \cdot \boldsymbol{s} + \boldsymbol{n},
\end{equation}
where $\boldsymbol{s}$ is the transmitted symbol vector, $\boldsymbol{n} \sim \mathcal{CN}(0, \sigma^2 \boldsymbol{I})$ is additive Gaussian noise with variance determined by the SNR, and $h$ is the channel fading coefficient. Specifically, $h=1$ for AWGN, $h \sim \mathcal{CN}(0,1)$ for Rayleigh fading, and $h = \sqrt{{K_R}/({K_R+1})} \, h_{\mathrm{LoS}} + \sqrt{{1}/({K_R+1})} \, h_{\mathrm{NLoS}}$ for Rician fading with factor $K_R$. The channel layer is non-trainable and differentiable, enabling gradient back-propagation for joint end-to-end training following established deep JSCC practice~\cite{ref2}.

\subsection{Joint Source-Channel Decoder} 
After receiving the noisy signal, the receiver decodes it using a joint source-channel decoder. The specific process can be expressed as $\hat{\boldsymbol{x}}=R\left(\boldsymbol{y}, \boldsymbol{\varphi}_\beta\right)$, where $\hat{\boldsymbol{x}}$ is the reconstructed image, $R(\cdot)$ represents a joint source-channel decoding network, and $\boldsymbol{\varphi}_\beta$ is the set of parameters for the corresponding decoding network. After this step, the signal is restored to a form that the user can understand, or the image is reconstructed.

Based on the above discussion of the network, the basic framework and operation process of the semantic communication system are introduced. Semantic information extraction and image reconstruction are considered key to the success of this system. Therefore, in this paper, the average Mean Squared Error (MSE) between the original input image and the reconstructed image is used as the loss function, which is defined as,

\begin{equation}
	\label{Eq_2}
	MSE=\frac{1}{L} \sum_{i=1}^{L}\left(x_i-\hat{x}_i\right)^2,
\end{equation}
where $x_i$ and $\hat{x}_i$ denote the $i$-th pixel value of the original and reconstructed images, respectively, and $L = 3 \times H \times W$ is the total number of pixel values in the image. A smaller MSE indicates that the reconstructed image is closer to the original image, reflecting better reconstruction quality.

\section{Image Semantic Encoder and Decoder Network}
\label{sec_4}
In this section, we first present a federated edge architecture for Swin Transformer-based semantic communication that extracts hierarchical semantic features from images. The system utilizes federated learning to facilitate distributed model training across multiple edge devices while maintaining user privacy. This approach enables collaborative learning of semantic representations from diverse user data without requiring centralized data collection, thereby ensuring robust communication performance across varying channel conditions.

\subsection{Swin Transformer-based Semantic Communication Model}
\begin{figure*}[!t]
	\centering
	\includegraphics[width=6.1in]{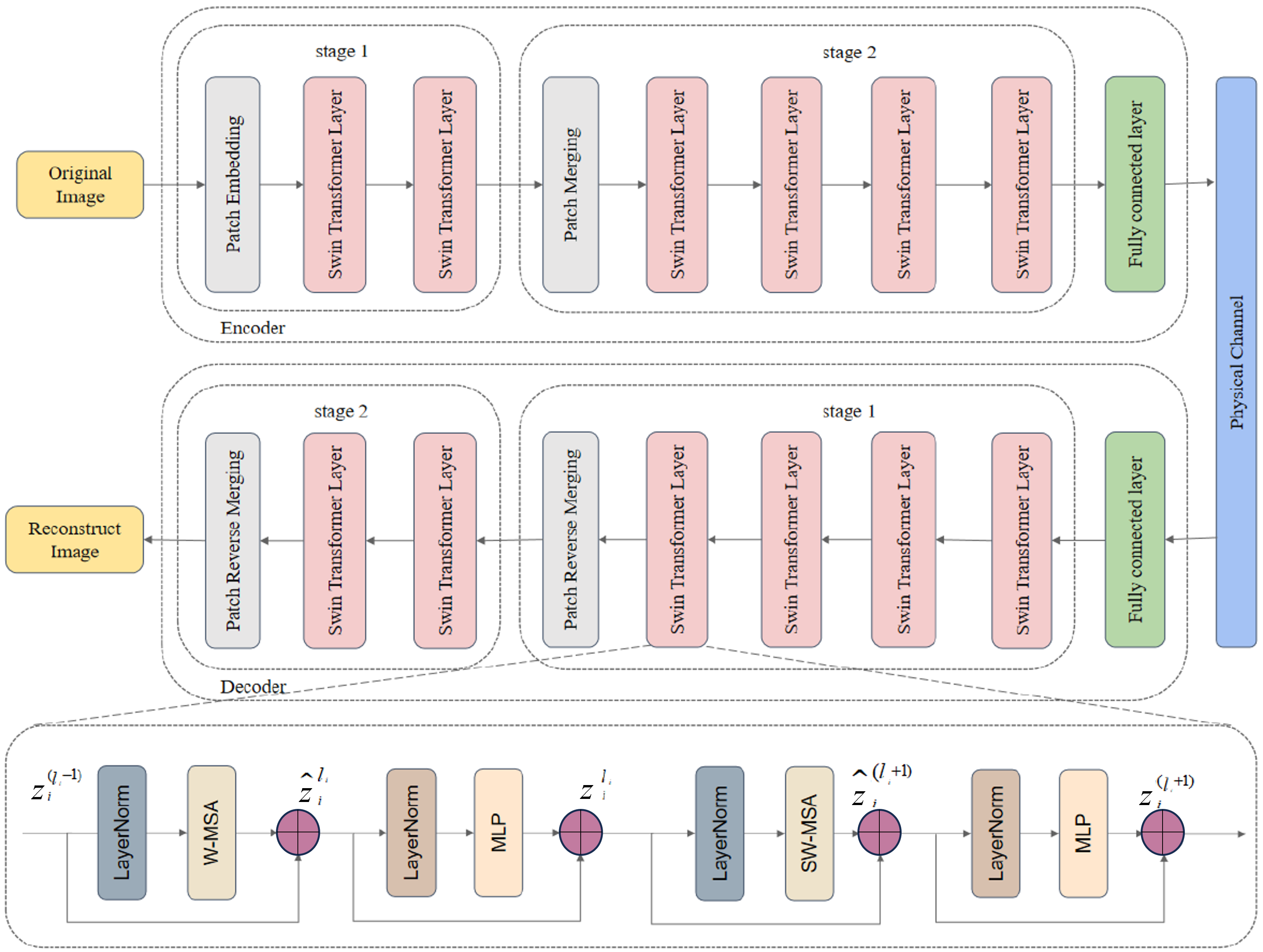}
	\caption{An illustration of the proposed semantic communication system structure.}
	\label{fig_3}
\end{figure*}

\textbf{Semantic Model.} 
Fig.\ref{fig_3} shows the semantic communication network structure based on Swin Transformer. In this paper, joint source-channel coding is adopted for semantic communication, utilizing the Swin Transformer module for both encoding and decoding. Additionally, a non-trainable fully connected layer is employed to simulate the physical channel.

At the transmitter, a set of training images 
$\mathbf{X}=\{\boldsymbol{x}_i\}$ is given, 
where $\boldsymbol{x}_i \in \mathbb{R}^{3 \times H \times W}$ 
denotes the $i$-th image, and $H$ and $W$ denote the height 
and width, respectively.
The images are first divided into non-overlapping $4{\times}4$ 
patches, reshaping the feature tensor to $(H/4,\,W/4,\,48)$.
The patch size is co-designed with the JSCC objective: it 
provides sufficient spatial granularity for semantic 
discrimination while keeping the token sequence length 
tractable for attention computation on resource-constrained 
UAV edge nodes.
Through the learnable embedding matrix $\boldsymbol{E}$, 
the feature blocks are projected into an embedding 
representation of dimension $C$, transforming the vector 
dimension to $(H/4,\,W/4,\,C)$; this completes Stage~1.

The feature blocks are then fed into Stage~2, where the 
patch merging layer concatenates each group of $2{\times}2$ 
adjacent patches, reducing the number of tokens to 
$H/8{\times}W/8$ while expanding the channel dimension 
to $4C$.
To perform downsampling, a $1{\times}1$ convolutional 
projection reduces the dimension from $4C$ to $2C$, 
after which the tokens pass through the Swin Transformer 
block for feature transformation.
The embedding dimension $C{=}32$ is co-designed with the 
UAV uplink bandwidth budget: the two-stage patch-merging 
pipeline concentrates multi-scale semantic information into 
a compact bottleneck of shape $(H/8,\,W/8,\,2C)$, yielding 
a compression ratio of $CR = (H/8{\times}W/8{\times}2C)\,/\, 
(3{\times}H{\times}W) = C/96 \approx 0.333$, 
suited for bandwidth-constrained UAV uplinks.

This hierarchical design addresses a fundamental limitation 
of alternative architectures for the UAV JSCC task.
Standard ViT~\cite{ref27} processes all patches at a fixed 
resolution with quadratic attention complexity 
$\mathcal{O}(n^2)$, which is prohibitive for edge inference 
on UAV platforms.
MobileViT~\cite{mehta2021mobilevit} reduces this cost by interleaving 
local convolutions with global self-attention, but remains 
constrained to a single feature scale, limiting its capacity 
to capture the multi-resolution spatial structures of UAV 
imagery under aggressive compression.
Swin Transformer resolves both issues simultaneously: 
window-restricted attention achieves linear complexity 
$\mathcal{O}(n)$, while the patch-merging pipeline 
produces the hierarchical multi-scale representations 
that prove critical for federated aggregation.

The output $\tilde{\boldsymbol{x}}$ passes through a fully connected 
layer to facilitate transmission in the channel, expressed as
\begin{equation}
	\label{Eq_3}
	\boldsymbol{s}=\boldsymbol{W}_1 \tilde{\boldsymbol{x}}+\boldsymbol{b}_1,
\end{equation}
where, $\boldsymbol{W}_1$ is the weight matrix of the full connection, and $ \boldsymbol{b}_1 $ is the bias term of the fully connected layer.

During training, the channel layer described in Section~\ref{sec:channel} is inserted between the encoder and decoder. For each batch, the fading coefficient $h$ and noise $\boldsymbol{n}$ are sampled according to the selected channel type and SNR. Since these operations are differentiable via the reparameterization trick, the encoder and decoder are jointly optimized end-to-end.

At the receiver, the decoder consists of the Swin Transformer blocks. The function is to decode the signal $ \boldsymbol{y} $ to restore the signal $ \hat{\boldsymbol{x}} $.

\textbf{Swin Transformer Layer}. 
In Swin Transformer, each Swin Transformer block \cite{ref28} consists of multiple Swin Transformer layers. Moreover, each Swin Transformer layer consists of two sub-layers: a Window-based Multi-Head Self-Attention (W-MSA) sub-layer and a Shifted Window Multi-Head Self-Attention (SW-MSA) sub-layer. The lower part of Fig.\ref{fig_3} illustrates the architecture of the Swin Transformer Layer. As shown in the figure, the input feature is first processed by the W-MSA sub-layer, which consists of LayerNorm, Multi-Head Self-Attention (MSA), and a residual connection, followed by a feed-forward sub-layer with LayerNorm, Multi-Layer Perceptron (MLP), and a residual connection. Subsequently, the output enters the SW-MSA sub-layer, which shares the same structure but replaces W-MSA with SW-MSA to enable cross-window information exchange. The input and output dimensions of the two sub-layers are identical, enabling direct sequential connection. The computation is formulated as follows, where the MLP contains two linear layers with the ReLU activation function.

The first sub-layer applies W-MSA:
\begin{equation}
	\label{Eq_4}
	\hat{\mathbf{z}}^l=\mathrm{W\text{-}MSA}\left(\mathrm{LN}\left(\mathbf{z}^{l-1}\right)\right)+\mathbf{z}^{l-1},
\end{equation}
\begin{equation}
	\label{Eq_5}
	\mathbf{z}^l=\operatorname{MLP}\left(\mathrm{LN}\left(\hat{\mathbf{z}}^l\right)\right)+\hat{\mathbf{z}}^l.
\end{equation}

The second sub-layer applies SW-MSA:
\begin{equation}
	\label{Eq_6}
	\hat{\mathbf{z}}^{l+1}=\mathrm{SW\text{-}MSA}\left(\mathrm{LN}\left(\mathbf{z}^l\right)\right)+\mathbf{z}^l,
\end{equation}
\begin{equation}
	\label{Eq_7}
	\mathbf{z}^{l+1}=\operatorname{MLP}\left(\mathrm{LN}\left(\hat{\mathbf{z}}^{l+1}\right)\right)+\hat{\mathbf{z}}^{l+1},
\end{equation}
where $\hat{\mathbf{z}}^l$ and $\mathbf{z}^l$ denote the intermediate output after the attention module and the final output after the MLP in the W-MSA sub-layer, respectively. Similarly, $\hat{\mathbf{z}}^{l+1}$ and $\mathbf{z}^{l+1}$ are the corresponding outputs of the SW-MSA sub-layer.

Specifically, MSA can be expressed as

\begin{equation}
	\label{Eq_8}
	\operatorname{MSA}(\mathbf{z})=\operatorname{Concat}\left(\boldsymbol{H}_1, \boldsymbol{H}_2, \cdots, \boldsymbol{H}_k\right) \boldsymbol{W}^{mal},
\end{equation}
where $\operatorname{Concat}(\cdot)$ denotes concatenation along the feature dimension, and $\boldsymbol{W}^{mal}$ is the output projection weight matrix. Each self-attention head $\boldsymbol{H}_i \, (i=1,2,\cdots,k)$ is computed as

\begin{equation}
	\label{Eq_9}
	\boldsymbol{H}_i=\operatorname{Attention}(\boldsymbol{Q}, \boldsymbol{K}, \boldsymbol{V})=\operatorname{SoftMax}\left(\boldsymbol{Q} \boldsymbol{K}^{\mathrm{T}} / \sqrt{d_k}\right) \boldsymbol{V},
\end{equation}
where $\boldsymbol{Q}$, $\boldsymbol{K}$, and $\boldsymbol{V}$ are the query, key, and value matrices obtained via linear projections from the input embeddings, respectively, and $d_k$ denotes the dimension of each key vector.

Unlike W-MSA, SW-MSA enables cross-window interaction by cyclically shifting the feature map by half a window size along both spatial dimensions, then computing self-attention within each resulting sub-window via a mask mechanism. The original spatial arrangement is restored by an inverse cyclic shift after attention computation. The attention in SW-MSA incorporates a relative position bias $\boldsymbol{B}$:

\begin{equation}
	\label{Eq_10}
	\operatorname{Attention}(\boldsymbol{Q}, \boldsymbol{K}, \boldsymbol{V})=\operatorname{SoftMax}\left(\boldsymbol{Q} \boldsymbol{K}^{\mathrm{T}} / \sqrt{d_k}+\boldsymbol{B}\right) \boldsymbol{V},
\end{equation}
where $\boldsymbol{B}$ is the learnable relative position bias matrix used in SW-MSA.

In the low-bandwidth regime of UAV image transmission, capturing semantic context across spatial regions is critical for maintaining reconstruction quality under aggressive compression. The alternating use of W-MSA and SW-MSA within each Swin Transformer block ensures that semantic context propagates across window boundaries without being confined to independent local regions, while maintaining linear computational complexity with respect to image size. This makes the architecture well-suited for real-time semantic processing on resource-constrained UAV edge hardware.

\subsection{Federated Learning Training}
The multi-user semantic communication system algorithm is developed by applying STSC to a FL framework, enabling accurate extraction of semantic information and efficient semantic communication while preserving user privacy. Since direct intervention in local computation is constrained by data privacy and security requirements, the focus is placed on parameter aggregation performed at the cloud server. While Fig.\ref{fig_1} illustrates the overall architecture of the algorithm, the specific training process is detailed below.

We use FedAvg \cite{ref29} as the aggregation algorithm for our federated training. Assume that $N_c$ clients participate in FL training. $ \boldsymbol{D}_k $  denotes the local Dataset of client $ E_k $, and $ \left|\boldsymbol{D}_k\right| $ is the size of the corresponding dataset. Thus, the training loss of the client based on the model in the local sample is

\begin{equation}
	\label{Eq_11}
	\operatorname{Loss}^k(\omega)=\frac{1}{\left|\boldsymbol{D}_k\right|} \sum_{i=1}^{\left|\boldsymbol{D}_k\right|} \operatorname{Loss}_i^k(\omega),
\end{equation}
where  $ \operatorname{Loss}^k(\cdot) $ is the training Loss of the local client $ E_k $, and $ \operatorname{Loss}_i^k(\cdot) $ is the loss of the sample $ \boldsymbol{x_i} $ corresponding to the client. In the proposed algorithm, $ \operatorname{Loss}(\cdot) $ is represented in \cref{Eq_2}.

Therefore, the global federated training loss can be obtained as a weighted average of each client's loss, where the weights are proportional to each client's dataset size. The specific formula is as follows.

\begin{equation}
	\label{Eq_12}
	\operatorname{Loss}^{\circ}(\omega)=\sum_{k=1}^{N_c} \frac{\left|\boldsymbol{D}_k\right|}{\left|\boldsymbol{D}_o\right|} \operatorname{Loss}^k(\omega),
\end{equation}
where $ \operatorname{Loss}^{\circ}(\omega) $ is the global loss after cloud aggregation, and $ |\boldsymbol{D}_o| $ is the global dataset size. The purpose of the FL training is to find the global optimal model that minimizes the sum of training losses over all data. The specific federated training process is as follows.

\textbf{Initialization.} 
The dataset is non-uniformly split and distributed across clients. The semantic communication network based on the Swin Transformer is initialized on a cloud server, and the global model parameters $ \boldsymbol{W}^o $ are broadcast to clients to build local STSC networks.

\textbf{Model training.} 
In the communication round $ t $, the client participating in the training will conduct local training on STSC using the local dataset. Moreover, the stochastic gradient descent method is used to update the local model parameters. For each batch $i$, the process can be expressed as,

\begin{equation}
	\label{Eq_13}
	\omega_{i+1}^k \leftarrow \omega_i^k-\eta \frac{\partial \operatorname{Loss}\left(\omega_i^k ; b\right)}{\partial \omega_i^k},
\end{equation}
where  $ \eta $ is the learning rate. Each client saves its own trained local parameters and uploads them to the cloud server.

\textbf{Parameter aggregation.} 
The cloud server aggregates model parameters from different clients and updates $ \boldsymbol{W}_{t}^o $ to $ \boldsymbol{W}_{t+1}^o $. The aggregation mechanism is shown in the following equation. Through this process, the global shared model can be updated.

\begin{equation}
	\label{Eq_14}
	w_{t+1} \leftarrow \sum_{k=1}^N \frac{\left|\boldsymbol{D}_k\right|}{\left|\boldsymbol{D}_o\right|} w_{t+1}^k,
\end{equation}
where $ N $ is the total number of clients participating in training, this aggregation mechanism provides a guarantee of system stability with respect to model convergence. Specifically, a certain degree of client-side fluctuations can be tolerated without affecting the final convergence of the model.

\begin{algorithm}[t!]
	\caption{Training Process}
	\begin{algorithmic}
		\STATE{Input:The number of clients $N_c$, the communication round $T$ and the datasets of client $\boldsymbol{D}_k$,$k=1...N_c$.} \\
		\STATE{Output:The trained global model.} \\
		\STATE{\textbf{Cloud Server executes:}} \\
		\STATE{Initialize the global network and the weight $\boldsymbol{W}^{\mathrm{o}}$;} \\
		\FOR{communication round $=1, \ldots, t, \ldots, T$.}
		\STATE At round $t$,the global model $\boldsymbol{W}_{t}^o$ is broadcasted to each client as $\boldsymbol{W}_{t}^k$. \\
		\FOR{each client  $k$ }
		\STATE $\boldsymbol{W}_{t+1}^k \leftarrow$ ClientUpdate $\left(\mathrm{k}, \boldsymbol{W}_t^k\right)$. \\
		\ENDFOR
		\STATE The cloud server aggregates each client's  $\boldsymbol{W}_{t+1}^k$ and updates $\boldsymbol{W}_{t}^o$  to $\boldsymbol{W}_{t+1}^o$ by \cref{Eq_14}. \\
		\STATE{$t$$\leftarrow$$t+1$.}\\
		\ENDFOR
		\STATE{\textbf{Clients update:}} \\
		\STATE{Download the local weight $\boldsymbol{W}^e$ and initialize the STSC network.} \\
		\STATE{Initialize the local training epoch $n_{\max}$.} \\
		\FOR{epoch $=1, \ldots, n, \ldots, n_{\max }$}
		\FOR{each batch $\left\{\boldsymbol{x}_i\right\}_{i=1}^{32}$ from $D_k$}
		\STATE Pass  through the local STSC network and update the network by the loss functions \cref{Eq_2}: \\
		\STATE {$\omega_{i+1}^{k}$ $\leftarrow$ $\omega_{i}^{k}-\eta \frac{\partial \text { Loss }\left(\omega_{i}^{k} ; b\right)}{\partial \omega_{i}^{k}}$.}\\
		\ENDFOR
		\STATE{$n$$\leftarrow$$n+1$.}\\
		\ENDFOR
	\end{algorithmic}
	\label{alg1}
\end{algorithm}

\noindent\textbf{Remark on the FL design.}
The adoption of FedAvg is motivated by the cross-silo 
federation setting considered in this work, where each 
client corresponds to either a UAV edge server or a 
trusted ground-side edge server handling training on 
behalf of lightweight terminal devices, as described 
in Section~\ref{sec_3}. In this architecture, UAV 
mobility affects only the intra-silo data collection 
phase, since parameter aggregation takes place between 
ground-side edge servers and the cloud rather than 
over the bandwidth-limited UAV uplink. The two 
processes operate in separate phases of the mission 
cycle and do not compete for the same channel resource, 
which justifies the ideal parameter transmission 
assumption adopted in Section~\ref{sec_5}.

\begin{table}[t!]
	\centering
	\caption{Hyper-parameter Settings.}
	\label{tab:hyperparams}
		\begin{tabularx}{\linewidth}{Xl}
			\toprule
			Hyper-Parameters               & Value  \\
			\midrule
			Number of Clients              & 3      \\
			Number of Communication Rounds & 60     \\ 
			Batch Size                     & 32     \\ 
			Learning Rate                  & 0.001  \\ 
			Embedding Dimension            & 32     \\ 
			\bottomrule
		\end{tabularx}
	\end{table}
	\textbf{Model convergence.} The parameters of the global model $ \boldsymbol{W}_{t+1}^o $  are downloaded to each client. Then, the next round of training commences. Repeat this step until a specified global training cycle is reached or the model converges. In particular, the proposed algorithm is summarized in Algorithm \ref{alg1}.
	
	\subsection{Privacy Analysis}
	\label{sec:privacy}
	This subsection analyzes the privacy guarantees of the proposed framework by formalizing the threat model and examining resilience against known attack vectors in both the training and inference phases.
	
	We consider two categories of adversaries: an honest-but-curious cloud server that faithfully executes FedAvg but may attempt to infer private information from received model updates $\boldsymbol{W}_k^{t+1}$, and a passive eavesdropper on the uplink channel who intercepts the encoded semantic features $\boldsymbol{y}$ during inference and attempts to reconstruct the original image.
	
	During the training phase, gradient inversion attacks~\cite{ref_DLG} represent the primary privacy threat, wherein an adversary reconstructs private training images by optimizing a dummy input whose gradient matches the observed model update. The proposed STSC framework mitigates this threat through three structural properties. First, the Swin Transformer architecture employs LayerNorm rather than BatchNorm, removing the batch-level statistical dependencies that gradient inversion methods typically exploit~\cite{ref_GradViT}, this makes gradient matching substantially harder than for BatchNorm-based CNNs. Second, batch-level training with a batch size of 32 forces the attacker to simultaneously recover all images from an averaged gradient, a combinatorially harder problem than single-image recovery~\cite{ref_DLG}. Third, each client performs $n_{\max}=1$ local epoch before uploading a single aggregated update, so the server observes a multi-step parameter difference rather than a single-batch gradient, further obfuscating individual sample contributions. We note that DLG is a baseline attack; more advanced methods such as GradViT~\cite{ref_GradViT} have shown improved reconstruction for Vision Transformers under favorable conditions, and resilience against stronger variants warrants further investigation.
	
	During the inference phase, the encoder compresses $\boldsymbol{x} \in \mathbb{R}^{3 \times H \times W}$ into a compact semantic representation with a compression ratio of approximately $1/3$, discarding pixel-level details in favor of task-relevant features. The subsequent channel corruption $\boldsymbol{y} = h \cdot \boldsymbol{s} + \boldsymbol{n}$ further degrades the fidelity of any inversion attempt. However, these provide passive rather than cryptographic privacy protection; recent work~\cite{ref_MIEA} has shown that intelligent eavesdroppers can reconstruct transmitted images with non-trivial quality under high-SNR conditions. Dedicated inference-phase defenses therefore remain necessary for privacy-critical deployments.
	
	For deployment scenarios requiring formal privacy guarantees, the proposed framework is readily compatible with differential privacy~\cite{ref_DP}, where calibrated noise is injected into $\boldsymbol{W}_k^{t+1}$ before uploading to provide $(\epsilon, \delta)$-guarantees, and with secure aggregation protocols~\cite{ref_SecAgg}, which prevent the server from inspecting individual client updates.
	
	\section{Experiment and Numerical Results}
	\label{sec_5}
	\subsection{Simulation Settings}
	
	In this paper, the CIFAR-10 dataset~\cite{ref30} is used for evaluation. This benchmark is adopted for its consistent use in representative deep JSCC works~\cite{ref2}, enabling direct comparison with existing methods, and the evaluation of semantic reconstruction quality under bandwidth constraints is fundamentally determined by the compression ratio rather than input resolution.
	
	The number of clients is set to $N_c=3$, each assigned a distinct and non-overlapping portion of the training data. This configuration reflects a realistic cross-silo federation scenario where a small number of UAV operational zones, each managed by a dedicated edge server, collaborate under a stable network topology. 
	Unlike cross-device FL with hundreds of mobile endpoints, cross-silo deployments typically involve 2--10 organizational participants~\cite{ref29}, and the FedAvg convergence guarantees established in~\cite{ref29} hold independently of the number of participating silos.
	The training and testing environment is Windows 11 with CUDA 12.1, and the deep learning framework is PyTorch 2.2.1.
	
	In the experiments, the process of downloading and uploading model parameters between the client and the cloud server is assumed to be ideal. The focus is on the effects of global models trained locally and aggregated via federated aggregation. To demonstrate the universality of the proposed method for semantic image networks, the classical Swin Transformer network is selected as the semantic feature extraction network, and the proposed semantic communication method is applied in subsequent experimental simulations. The following introduces the initialization parameter settings for the semantic communication network based on Swin Transformer. Firstly, the network hyperparameter $C$ is set to 32. All network training uses MSE as the loss function, with 60 communication rounds, a batch size of 32, a learning rate of 0.001, and the Adam optimizer \cite {ref31}. In addition, to verify the robustness of the semantic communication system architecture under various complex channel conditions, the channel types can be selected as Gaussian, Rician, and Rayleigh channels. In all tests, the image Compression Ratio (CR) is set to 0.333. The specific training parameter Settings are shown in Table \ref{tab:hyperparams}.
	
	\begin{figure*}[t!]
		\centering
		\subfloat[AWGN channel]{
			\includegraphics[width=0.32\textwidth]{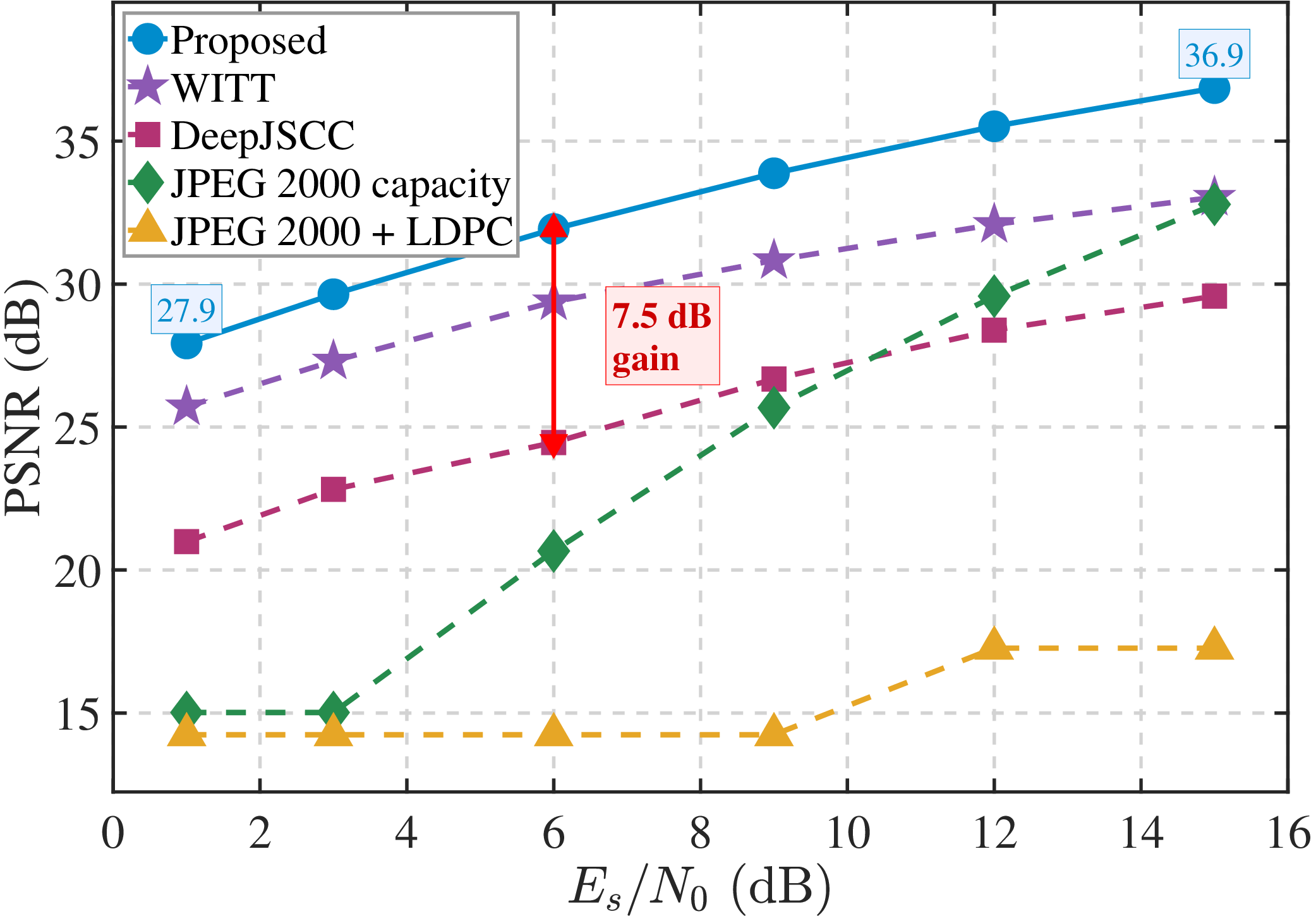}
			\label{fig4a}}
		\subfloat[Rician channel (K=10)]{
			\includegraphics[width=0.32\textwidth]{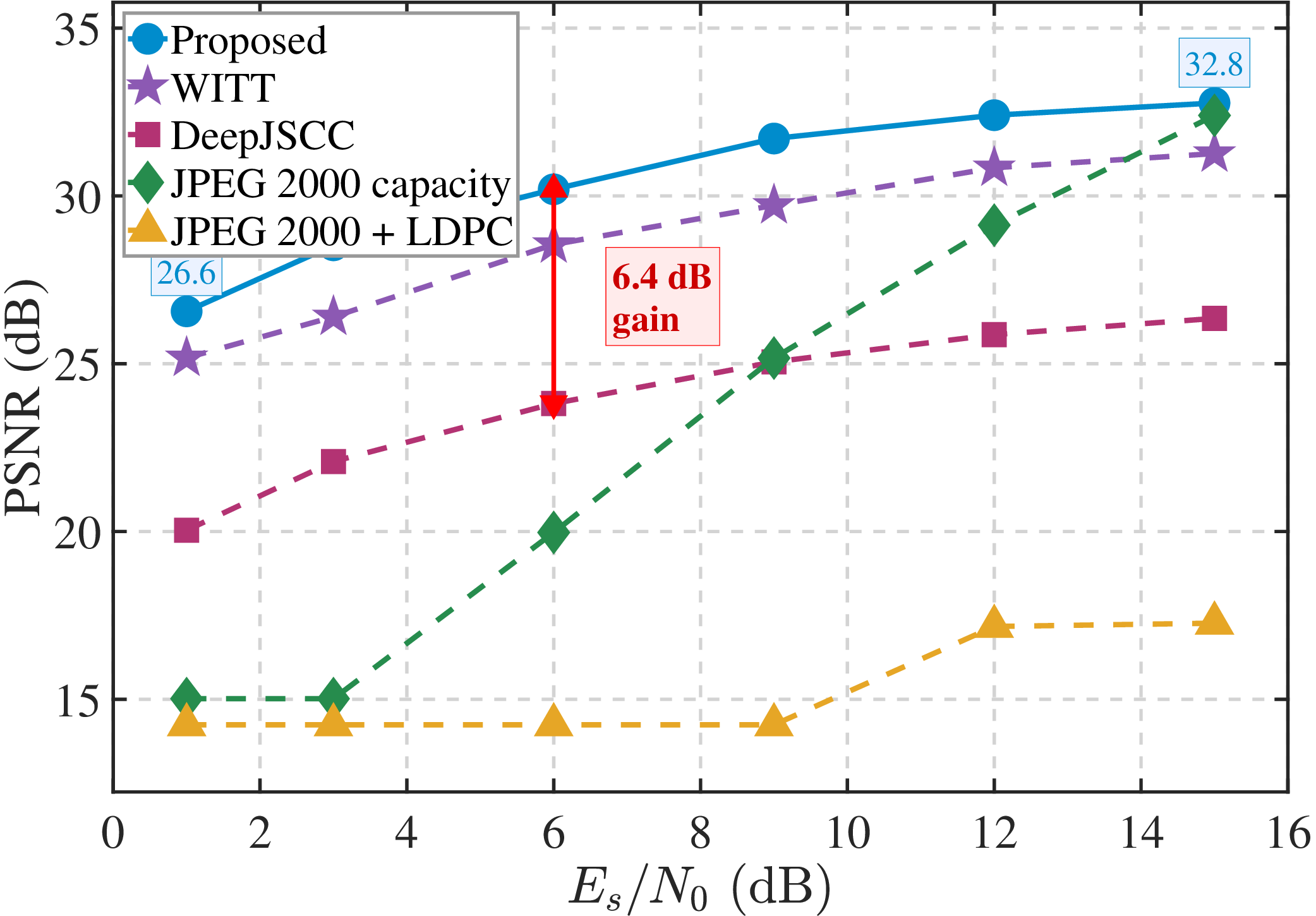}
			\label{fig4c}}
		\subfloat[Rayleigh channel]{
			\includegraphics[width=0.32\textwidth]{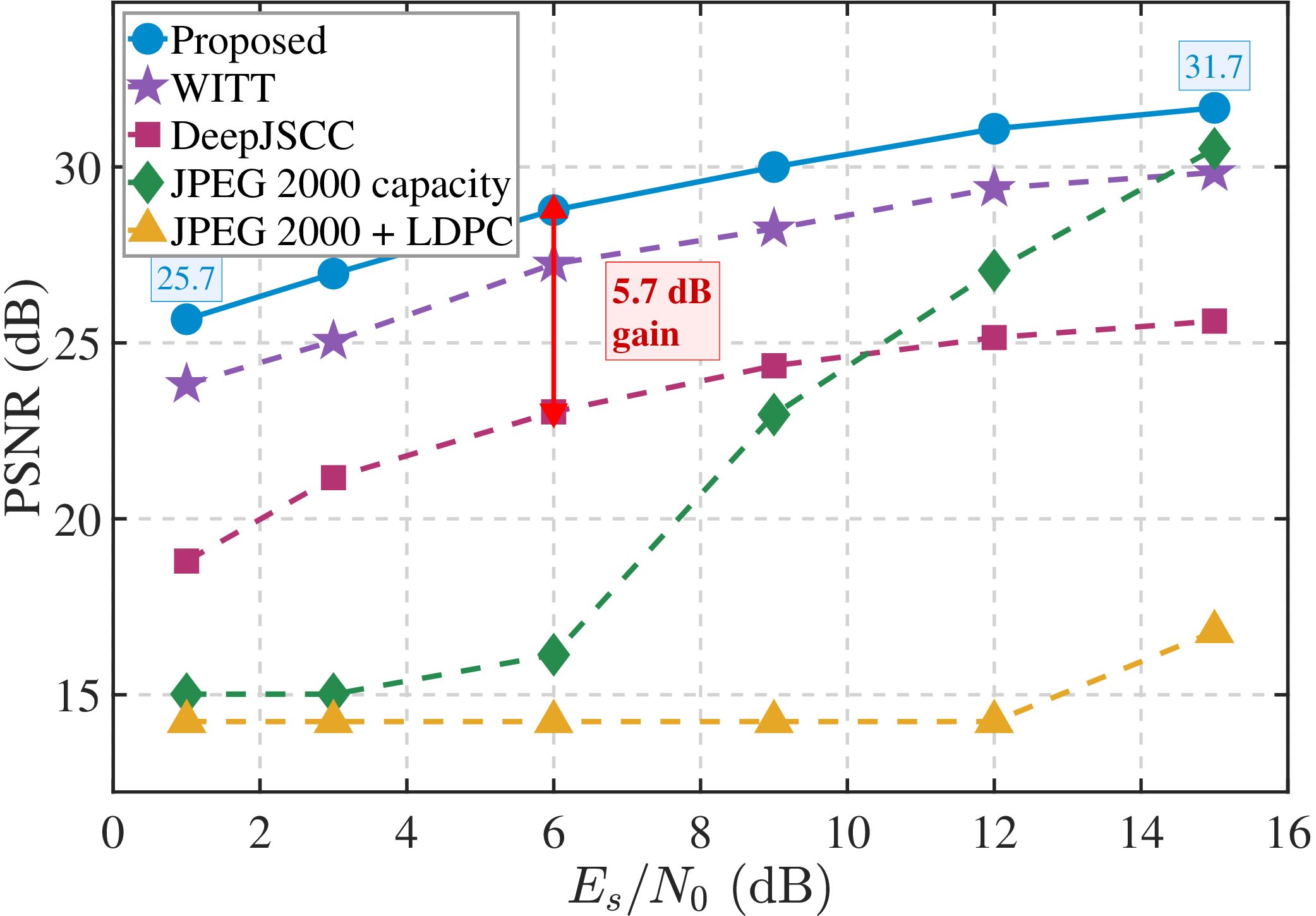}
			\label{fig4e}}
		\caption{PSNR performance comparison under different channel conditions.}
		\label{fig_4} 
	\end{figure*}
	
	To validate the effectiveness of the proposed method, we conduct comprehensive comparisons with both deep learning-based and 
	traditional communication schemes across three representative channel conditions: AWGN, Rician fading ($K_R=10$ dB), and Rayleigh fading. These channels span the typical air-to-ground propagation scenarios for UAV communications, from strong line-of-sight conditions to severe multipath fading. The Rician factor $K_R=10$ dB is adopted to represent typical medium-altitude UAV communication scenarios. Specifically, we select the representative DeepJSCC~\cite{ref2} and WITT~\cite{ref19} as the deep learning baselines. For traditional source-channel separation schemes, we employ JPEG 2000 as the source codec, LDPC for channel coding, and 16-QAM modulation, denoted as JPEG 2000+LDPC. For visual-quality comparison experiments, we use JPEG 4-QAM to ensure reliable transmission at low SNR. Additionally, we include the theoretical channel capacity limit of JPEG 2000 compression, denoted as JPEG 2000 capacity, which serves as the performance upper bound for conventional digital communication systems.
	
	\subsection{Simulation Results}
	\subsubsection{Reconstruction Quality Comparison}
	To comprehensively evaluate the effectiveness of semantic feature extraction and transmission, we conduct experiments comparing the proposed method against DeepJSCC, WITT, and conventional separation-based schemes (JPEG 2000 with capacity-achieving codes and JPEG 2000 + LDPC) under three distinct channel conditions. 
	As shown in Fig.~\ref{fig_4}, the proposed method consistently outperforms all baseline approaches across the entire SNR range. At $SNR = 6$~dB, the proposed method achieves PSNR gains of 7.5~dB, 6.4~dB, and 5.7~dB over DeepJSCC under AWGN, Rician, and Rayleigh channels, respectively. This improvement is attributed to the hierarchical multi-scale feature extraction capability of the Swin Transformer architecture, which captures task-relevant semantic information more effectively than the convolutional backbone of DeepJSCC during joint source-channel encoding.
	
	The separation-based schemes exhibit the well-known \textit{cliff effect}: PSNR remains severely degraded at low SNR due to decoding failures, while saturating at high SNR due to inherent lossy compression. In contrast, both learned JSCC methods exhibit graceful degradation, with PSNR increasing monotonically with channel quality. The proposed method maintains a significant advantage even at $SNR = 0$~dB, achieving 27.9~dB, 26.6~dB, and 25.7~dB under AWGN, Rician, and Rayleigh channels, respectively.
	
	Since the locally trained STSC model without federated aggregation is architecturally equivalent to WITT~\cite{ref19} under the same configuration, the comparison between the federated global model and local models in Fig.~\ref{fig_7} directly quantifies the gain over WITT. The federated global model consistently outperforms all local models by 2--3~dB across the entire SNR range under all three channel conditions, demonstrating that federated aggregation provides substantial reconstruction quality improvement beyond what the Swin Transformer architecture alone can achieve.
	
	\subsubsection{Federated Learning Performance Evaluation}
\begin{figure*}[!t]
		\centering
		\subfloat[AWGN channel]{
			\includegraphics[width=0.32\textwidth]{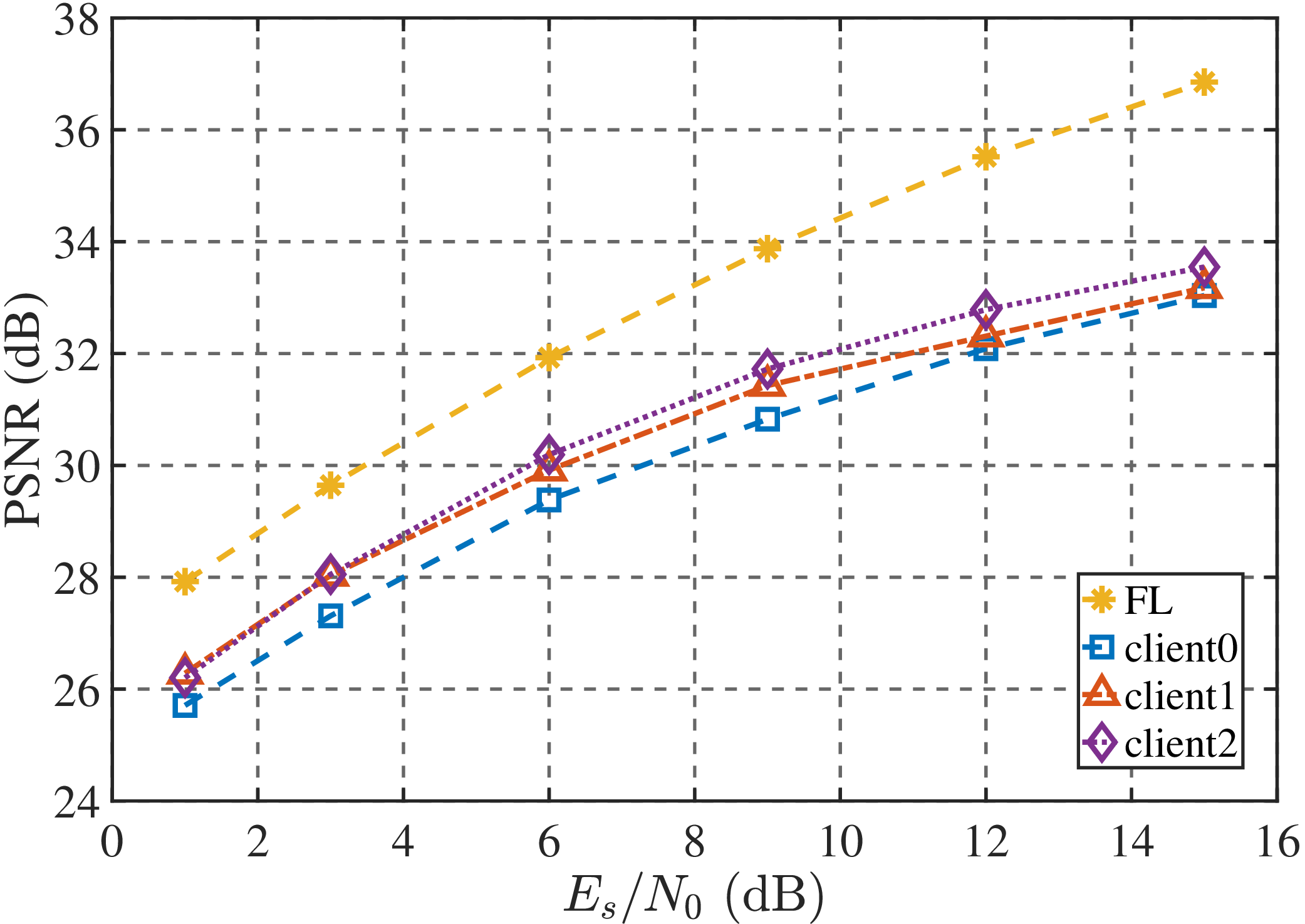}}
		\subfloat[Rician channel ($K_R$=10)]{
			\includegraphics[width=0.32\textwidth]{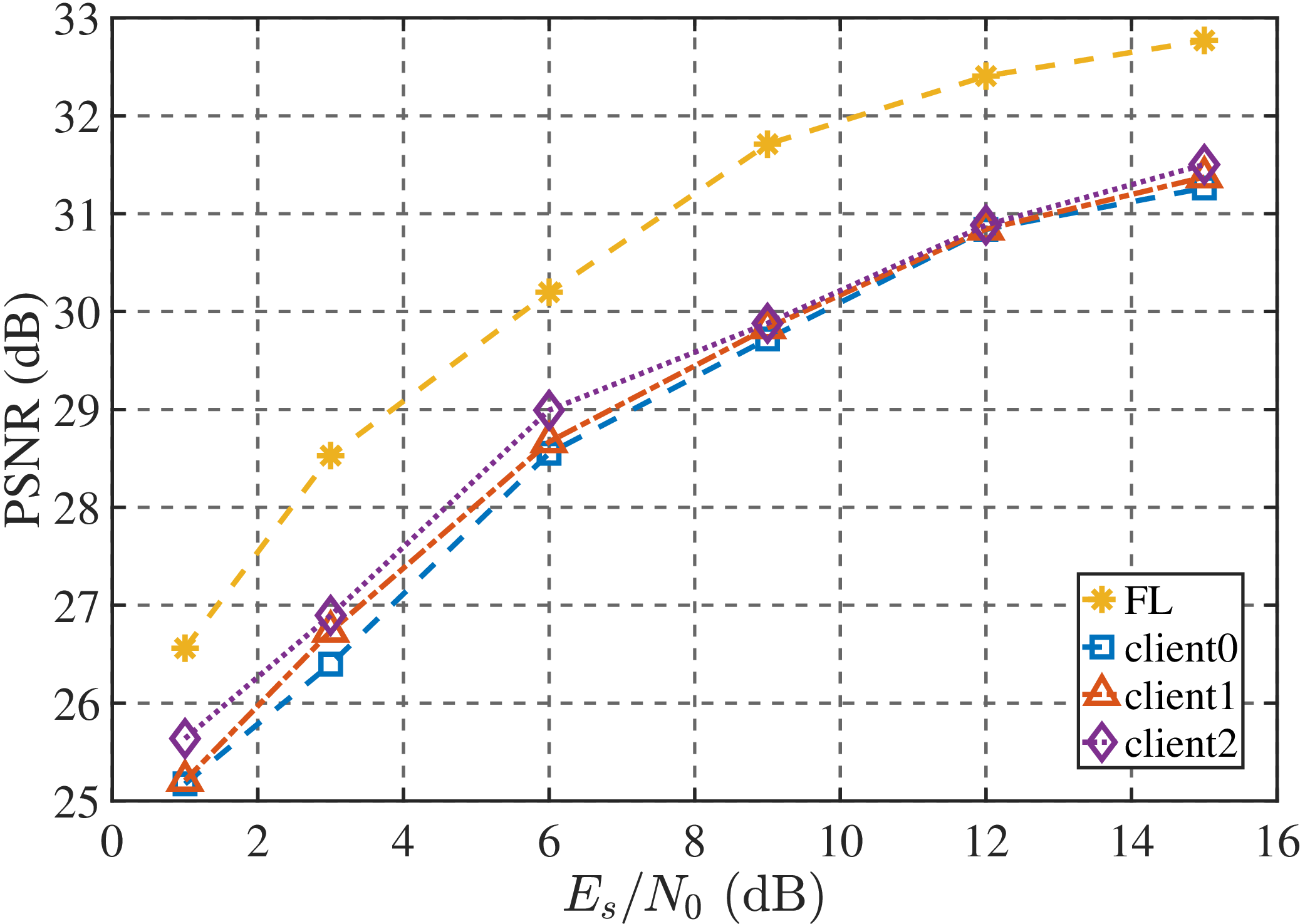}}
		\subfloat[Rayleigh channel]{
			\includegraphics[width=0.32\textwidth]{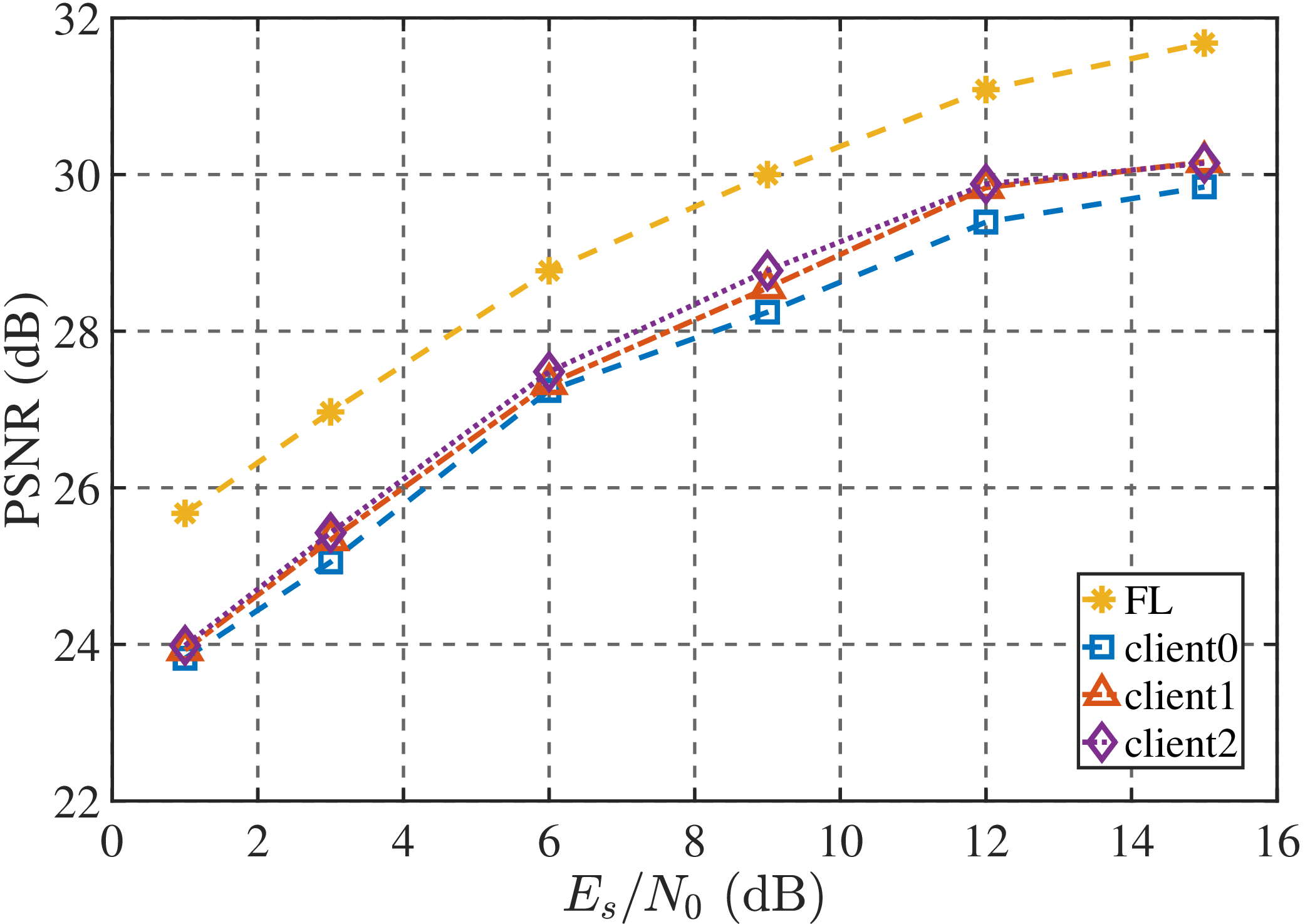}}
		\caption{PSNR versus $E_s/N_0$ comparison between the global model and local models. The global model trained via federated learning achieves 2-3 dB PSNR gain over local models, demonstrating improved noise resilience and generalization through model aggregation.}
		\label{fig_7} 
	\end{figure*}

	Fig.~\ref{fig_7} illustrates the PSNR performance comparison between the global model obtained through federated learning aggregation and individual local models (client0, client1, client2) across three channels. The federated global model consistently outperforms all local models across the entire SNR range under all channel conditions. Specifically, the global model achieves approximately 2--3 dB PSNR gain over local models. For instance, under the AWGN channel, the global model attains 36.9 dB at $SNR  = 15$ dB, compared to approximately 33-34 dB achieved by local models. Similar performance gains are observed under Rician and Rayleigh fading channels.
	
	This performance enhancement is attributed to the federated aggregation mechanism, which enables the global model to implicitly learn from diverse data distributions across multiple clients, thereby acquiring more comprehensive semantic representations and improved generalization. Additionally, exposure to heterogeneous channel conditions during distributed training enhances robustness against various channel impairments. These results demonstrate that the proposed federated learning framework not only preserves data privacy by avoiding the sharing of raw data among clients but also achieves superior reconstruction quality through collaborative model optimization.
	
	\subsubsection{Convergence Analysis}
\begin{figure*}[!t]
		\centering
		\subfloat[AWGN channel]{
			\includegraphics[width=0.25\textwidth]{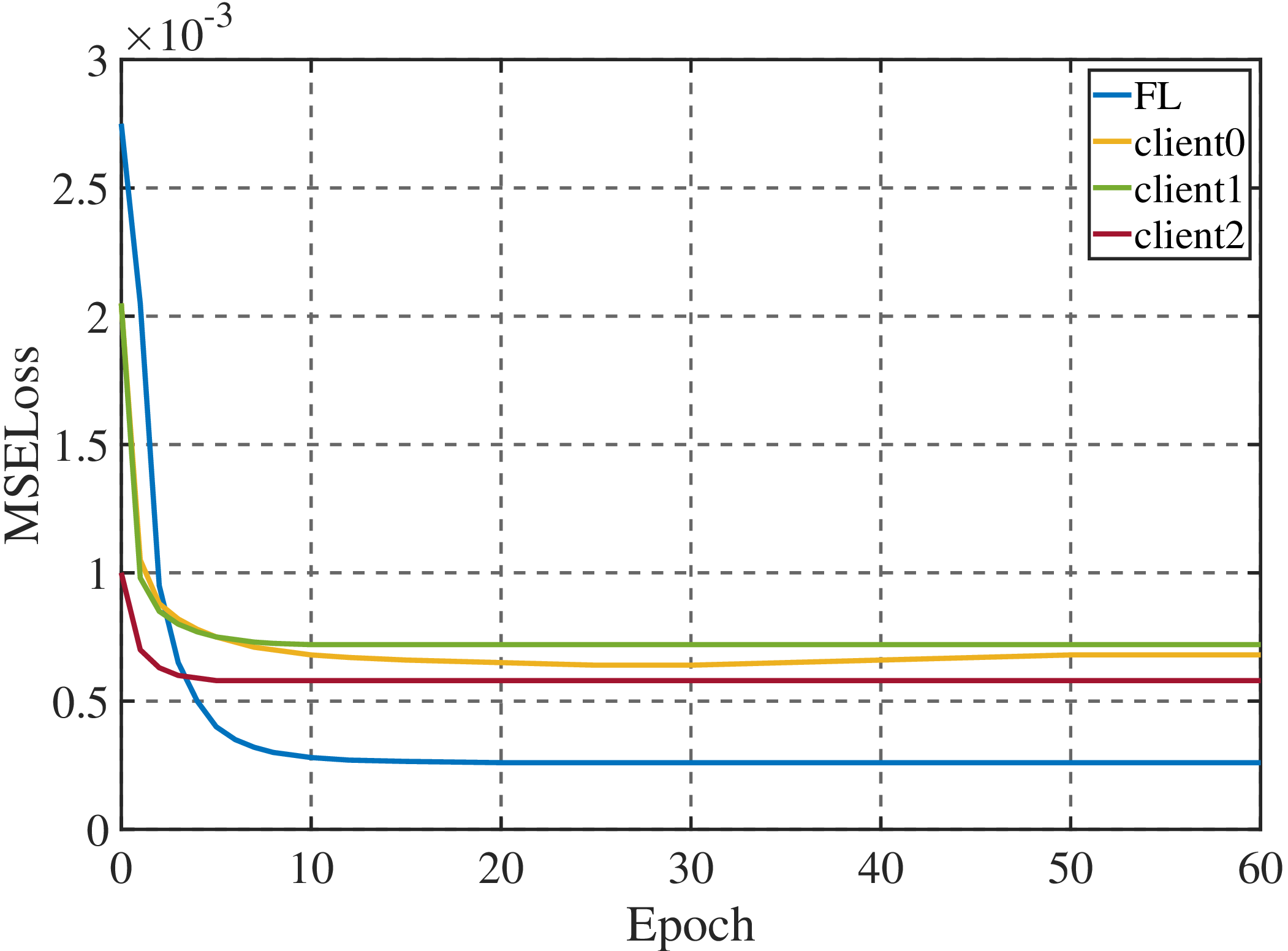}}
		\subfloat[Rayleigh channel]{
			\includegraphics[width=0.25\textwidth]{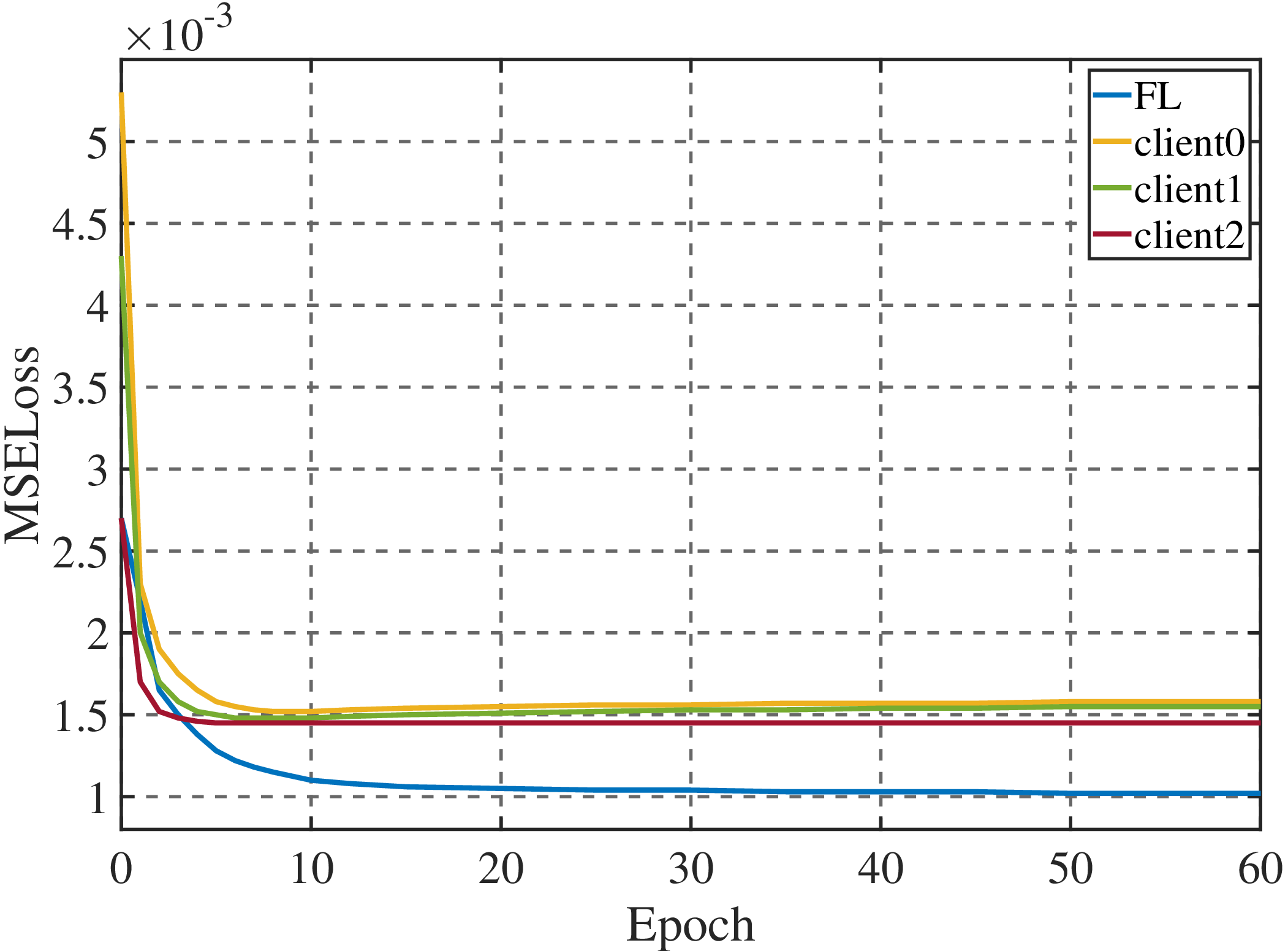}}
		\subfloat[AWGN channel]{
			\includegraphics[width=0.25\textwidth]{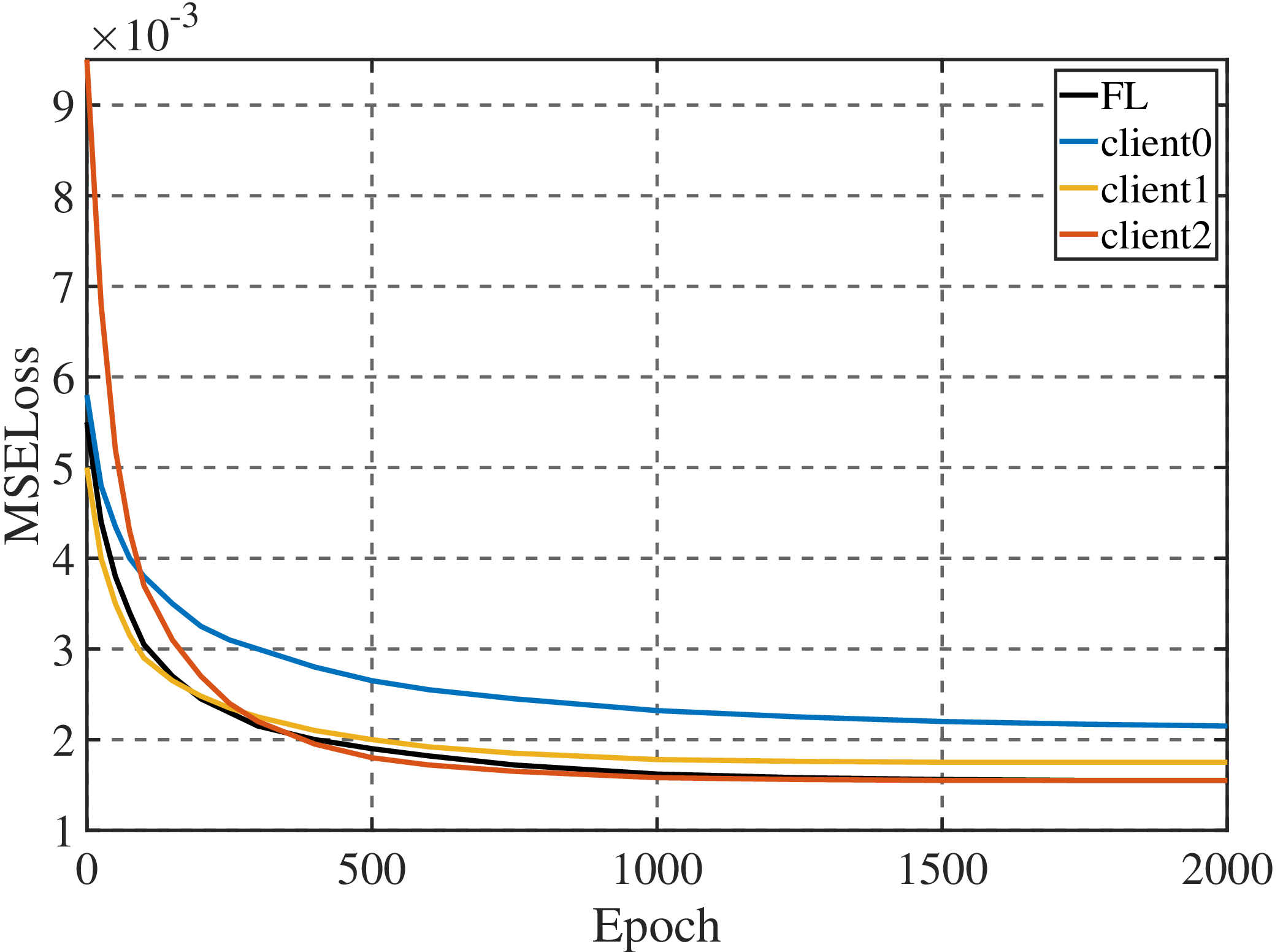}}
		\subfloat[Rayleigh channel]{
			\includegraphics[width=0.25\textwidth]{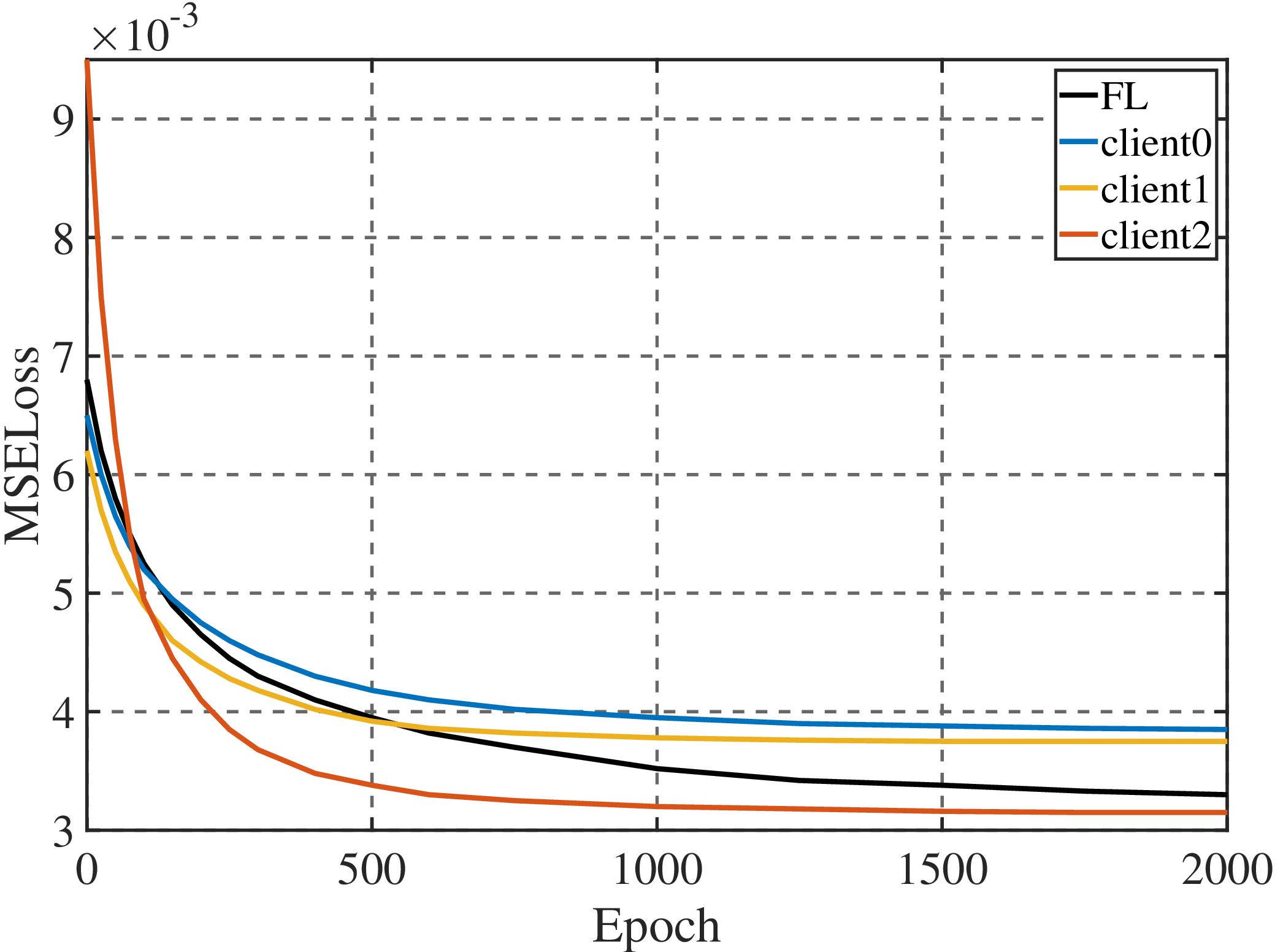}}
		\caption{Convergence results of the training MSE loss versus communication rounds (epochs), with the global and local models based on STSC at the left and DeepJSCC at the right. Compared to the global model of DeepJSCC, the federated convergence of the STSC model achieves a significantly faster processing speed and lower loss accuracy.}
		\label{fig_6} 
	\end{figure*}

	To evaluate the training dynamics of the proposed STSC and DeepJSCC models within the FL framework, we conduct convergence analysis at $SNR = 12$ dB. Fig.~\ref{fig_6} illustrates the MSE loss evolution over training epochs for both federated global models and local client models under AWGN and Rayleigh channels.
	
	As shown in Fig.~\ref{fig_6}(a) and (b), the STSC-based models exhibit rapid convergence within 60 epochs. The loss decreases sharply during the initial training phase and stabilizes thereafter. Notably, the federated global model consistently achieves lower converged loss compared to all local models, demonstrating that federated aggregation effectively leverages distributed data to learn more comprehensive semantic representations without direct data sharing.
	
	In contrast, as depicted in Fig.~\ref{fig_6}(c) and (d), DeepJSCC-based models require approximately 2000 epochs to converge. Additionally, the federated global model achieves only marginal improvement over local models, suggesting that the simple convolutional architecture of DeepJSCC limits its capacity to benefit from federated aggregation.
	
	\subsubsection{Visual Reconstruction Quality}
\begin{figure*} [t!]
		\centering
		\subfloat[AWGN channel]{
			\includegraphics[width=5in]{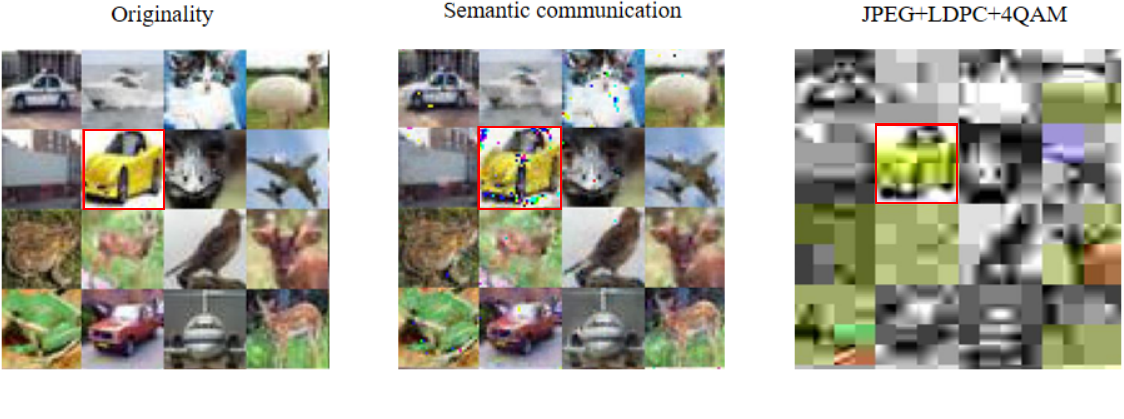}
			\label{5a}}
		\\
		\subfloat[Rayleigh channel]{
			\includegraphics[width=5in]{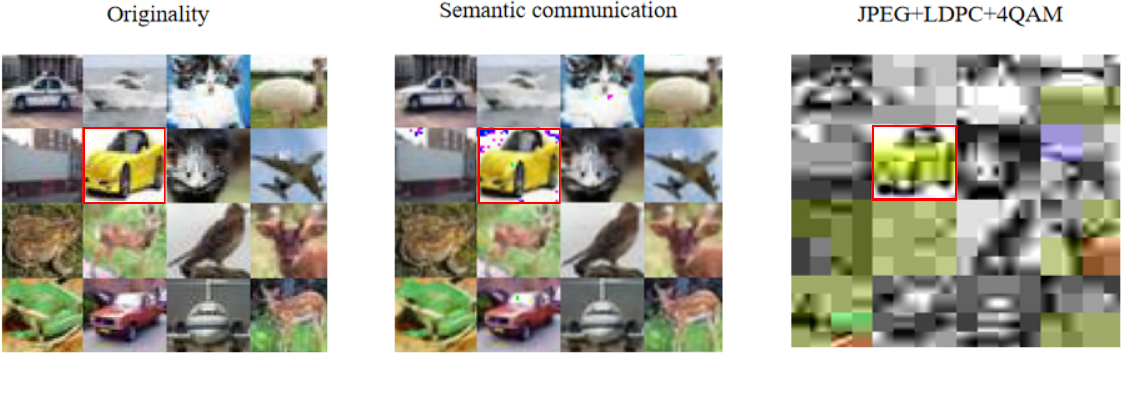}
			\label{5b}}
		\caption{Comparison of reconstructed images obtained by different methods. The STSC model demonstrates superiority over traditional communication algorithms by visually reconstructing images with higher fidelity.}
		\label{fig_5} 
	\end{figure*}

	To further demonstrate the superiority of the STSC model over the traditional communication algorithm, this section presents a visual reconstruction of the image. Taking $SNR=15$ as an example, the results in Fig.\ref{fig_5} (right) show that most of the reconstructed JPEG+LDPC+QAM images display only blocks of pixels, compared to the randomly selected original image in Fig.\ref{fig_5} (left). However, in Fig.\ref{fig_5} (middle), although some pixels experience transmission errors, the proposed algorithm maintains the image's overall visual quality and produces a more visually appealing result.
	\begin{figure}[h]
		\centering
		\includegraphics[width=0.4\textwidth]{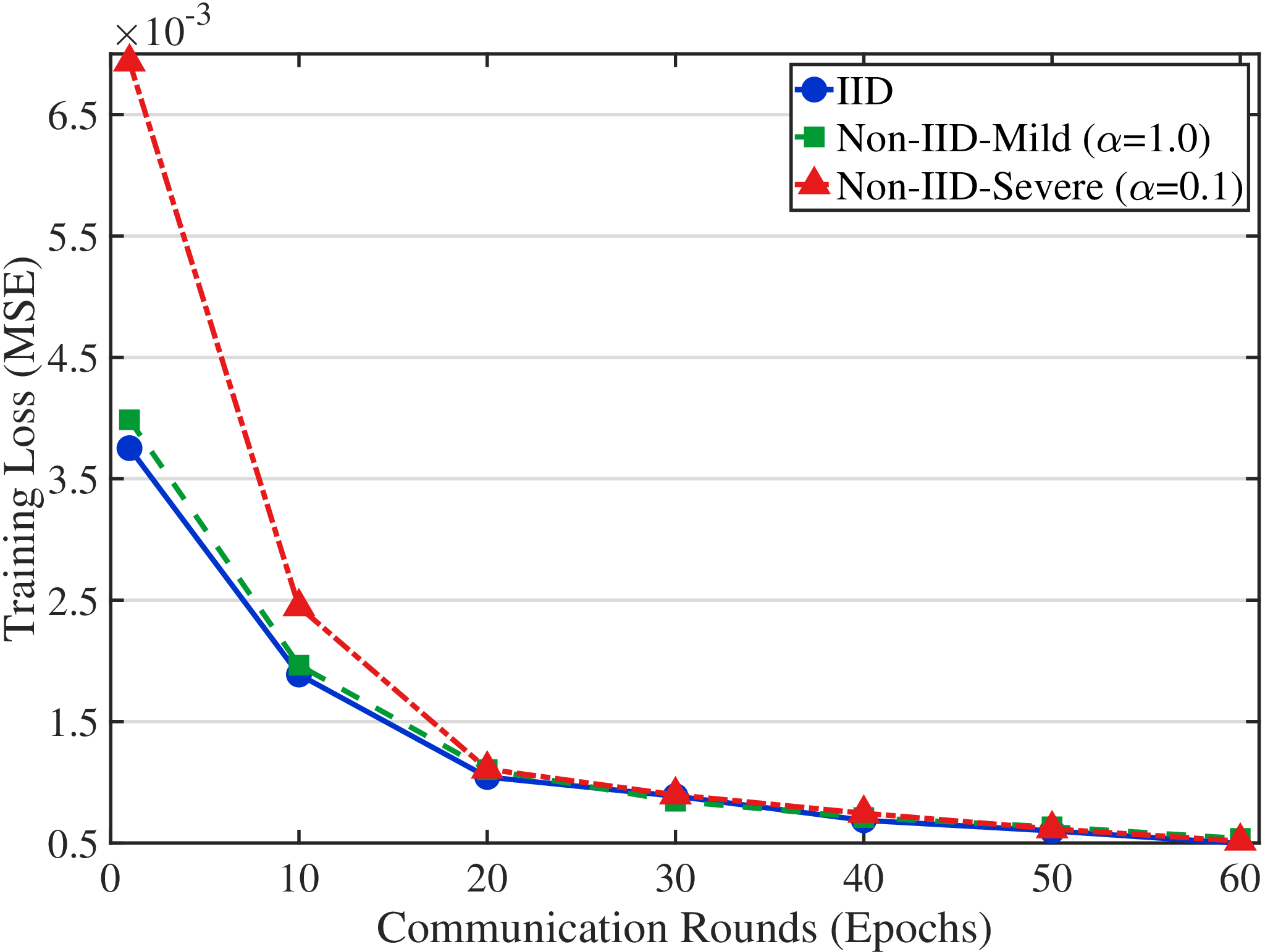}
		\caption{Convergence comparison under IID and Non-IID data distributions on Rician channel ($K_R$=10@SNR=12dB). The global FL model maintains stable convergence even under severe Non-IID conditions.}
		\label{fig:noniid}
	\end{figure}
	
	\subsubsection{Robustness under Non-IID Data Distributions}
	Real-world UAV deployments often encounter heterogeneous data distributions due to geographic diversity and varying mission scenarios. To evaluate the robustness of our federated learning framework, we conduct experiments across three data distribution settings: IID (uniform distribution), Non-IID-Mild ($\alpha=1.0$), and Non-IID-Severe ($\alpha=0.1$), where $\alpha$ denotes the concentration parameter of the Dirichlet distribution. Specifically, for each class $c$, a probability vector $\boldsymbol{p}_c \sim \mathrm{Dir}(\alpha \cdot \boldsymbol{1}_{N_c})$ is sampled, and the training samples of class $c$ are distributed among $N_c$ clients according to $\boldsymbol{p}_c$. A smaller $\alpha$ yields more skewed distributions: when $\alpha=1.0$, clients receive moderately uneven class proportions; when $\alpha=0.1$, most of each client's local data concentrates on only one or two classes, simulating the scenario where UAVs operating in different regions capture highly specialized imagery.
	
	Fig.~\ref{fig:noniid} presents the training loss convergence under different data distributions on the Rician channel ($K_R=10$) at $SNR = 12$ dB. All three settings converge to comparable loss values within 60 communication rounds. The Non-IID-Severe case exhibits slightly higher loss during the initial training phase, but rapidly aligns with the IID baseline thereafter. The negligible performance gap across different levels of data heterogeneity indicates that the hierarchical semantic representations learned by the Swin Transformer are inherently robust to local data bias, which is favorable for practical UAV swarm scenarios.
	
	We further note that the Dirichlet-based partitioning naturally introduces non-uniform data quantities across clients: under $\alpha=0.1$, per-client dataset sizes vary by up to a factor of three. Since the FedAvg aggregation in Eq.~(\ref{Eq_14}) weights each client's contribution proportionally to $|D_k|/|D_o|$, the framework inherently accommodates heterogeneous data volumes without requiring explicit balancing. Regarding computational heterogeneity, the cross-silo architecture in Section~\ref{sec_3} assigns STSC training to dedicated edge servers with uniform processing capacity, which is consistent with standard cross-silo FL assumptions~\cite{ref22}.
	
\begin{figure}[t]
		\centering
		\includegraphics[width=0.44\textwidth]{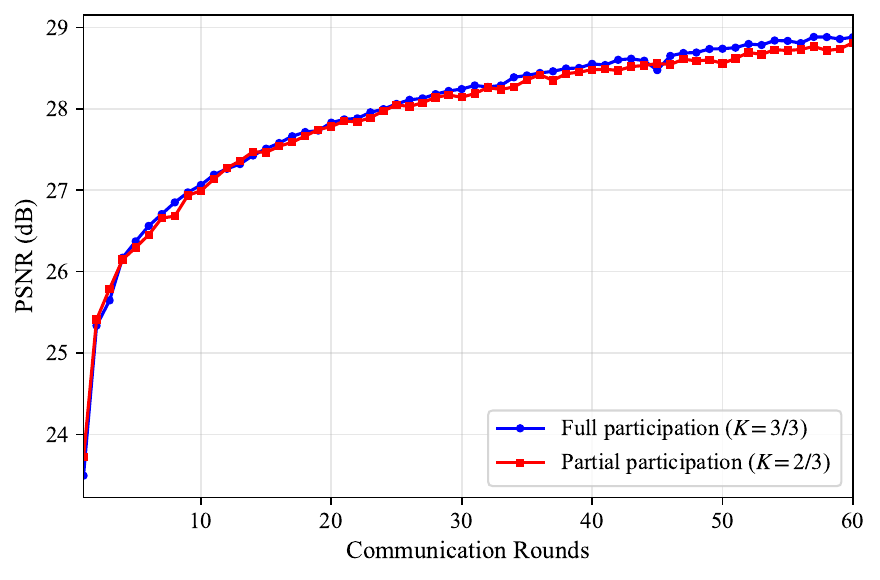}
		\caption{Convergence comparison under full participation 
			($\mathcal{K}=3/3$) and partial participation ($\mathcal{K}=2/3$) on the Rician 
			channel ($K_R=10$) at $SNR=3$~dB.}
		\label{fig:partial}
	\end{figure}

	To evaluate robustness under partial client participation, we conduct experiments where only $\mathcal{K}=2$ out of $N_c=3$ clients are randomly selected in each communication round, simulating scenarios where UAVs are temporarily unavailable due to mission duty cycles or connectivity issues. Fig.~\ref{fig:partial} compares the convergence behavior under full participation ($\mathcal{K}=3$) and partial participation ($\mathcal{K}=2$) on the Rician channel. Both settings converge to comparable PSNR values within 60 communication rounds, with the partial participation case achieving 28.8~dB compared to 28.9~dB under full participation. The negligible 0.1~dB gap confirms that the FedAvg aggregation mechanism produces stable global updates even when a subset of clients contributes in each round, demonstrating the framework's suitability for practical UAV deployments with intermittent connectivity.
	
	\subsubsection{Privacy Evaluation}
	\label{sec:privacy_exp}
	
	To empirically validate the privacy analysis in Section~\ref{sec:privacy}, we evaluate both training-phase and inference-phase privacy under the Rician channel.
	
\begin{figure}[t]
		\centering
		\includegraphics[width=\linewidth]{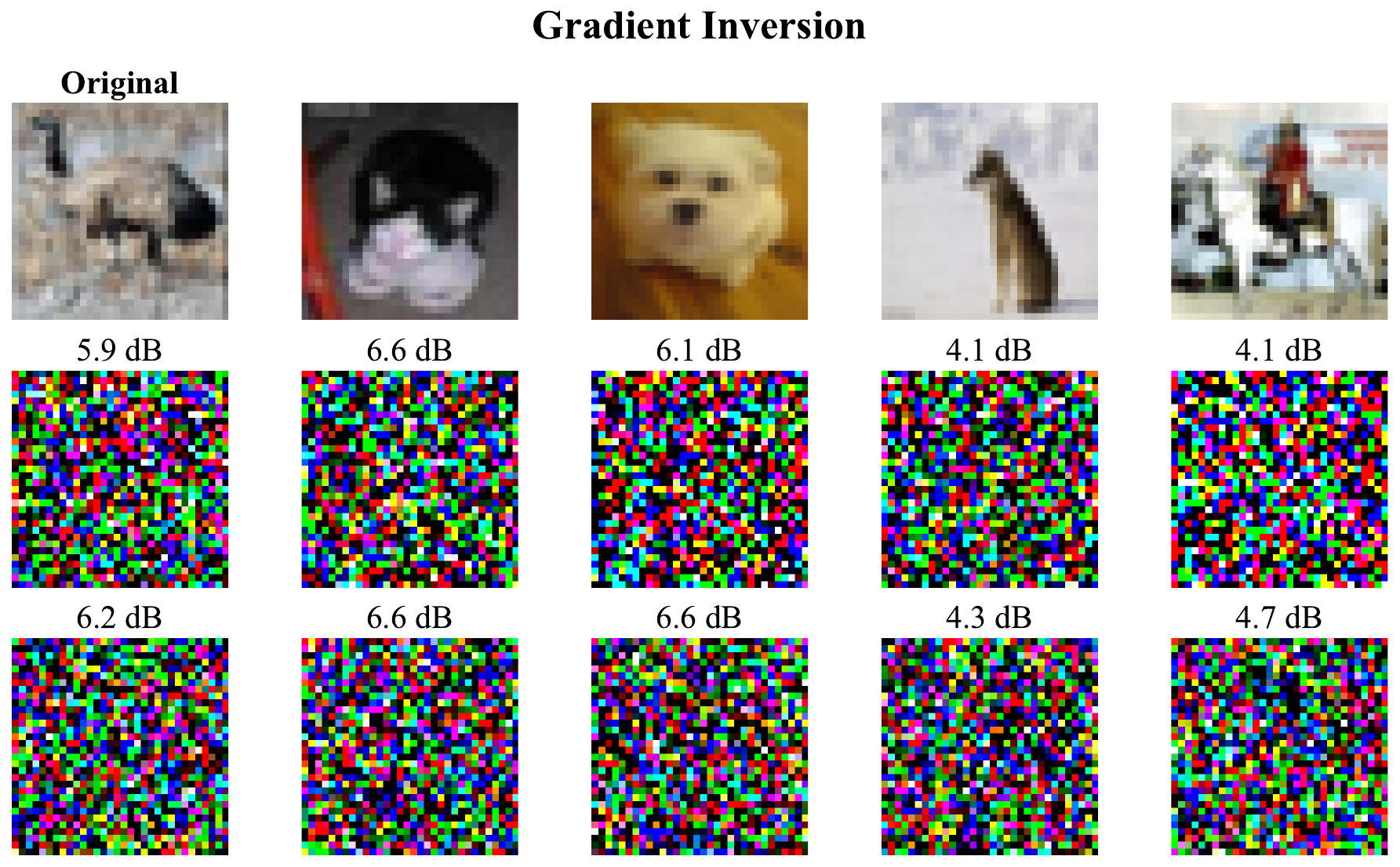}
		\caption{DLG gradient inversion results under the Rician channel ($K_R=10$, $SNR=3$~dB). Top: original images. Middle: batch size $=1$. Bottom: batch size $=32$. All reconstructions are indistinguishable from random noise.}
		\label{fig:gradient_attack}
	\end{figure}

	For the training phase, we apply the Deep Leakage from Gradients (DLG) method~\cite{ref_DLG} as a baseline gradient inversion attack, which optimizes a randomly initialized dummy image over 300 L-BFGS iterations to match the gradient computed from the true input. We test both single-image gradients (batch size $=1$, the most favorable setting for the attacker) and realistic batch gradients (batch size $=32$). As shown in Fig.~\ref{fig:gradient_attack}, both settings yield reconstructions that are visually indistinguishable from random noise, achieving only 5.3~dB / 5.7~dB PSNR and near-zero SSIM (0.003 / 0.005), respectively. Compared with the legitimate decoder's 28.4~dB PSNR, the quality gap exceeds 22~dB. This resistance is primarily attributed to the exclusive use of LayerNorm, which eliminates the batch-level statistical leakage exploited by attacks designed for BatchNorm-based architectures~\cite{ref_GradViT}.
	
\begin{figure}[t]
		\centering
		\includegraphics[width=\linewidth]{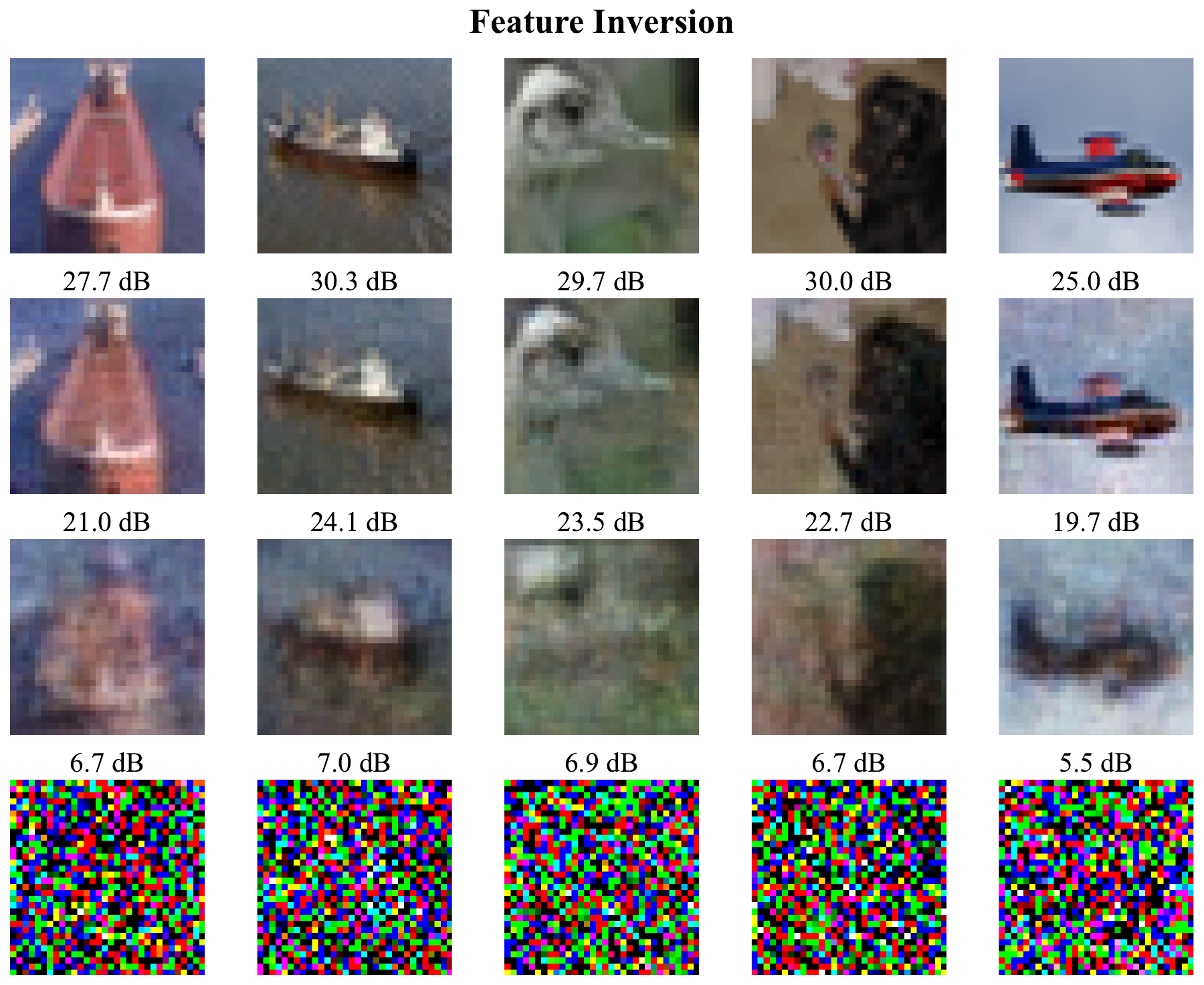}
		\caption{Semantic feature inversion results under the Rician channel ($K=10$, $SNR=3$~dB). From top to bottom: original, legitimate decoder, trained inversion network, and optimization-based inversion.}
		\label{fig:feature_inversion}
	\end{figure}

	For the inference phase, we simulate a passive eavesdropper who intercepts the channel-corrupted semantic features $\boldsymbol{y}$ and attempts image reconstruction via three approaches: the legitimate decoder (upper bound), a separately trained inversion network with access to 500 known feature-image pairs, and optimization-based inversion without any training data. Table~\ref{tab:feature_inversion} and Fig.~\ref{fig:feature_inversion} summarize the results. The inversion network achieves 22.2~dB PSNR and 0.629 SSIM, representing a 6.2~dB degradation from the legitimate decoder despite requiring a substantial corpus of known pairs that is difficult to obtain in practice. The optimization-based approach, which reflects the most realistic attack scenario, yields only 6.8~dB PSNR and 0.043 SSIM, demonstrating that the lossy semantic compression combined with channel noise renders intercepted features practically non-invertible.
	
	\begin{table}[t]
		\centering
		\caption{Inference-phase privacy evaluation under the Rician channel ($K=10$, $SNR=3$~dB).}
		\label{tab:feature_inversion}
		\begin{tabular}{lcc}
			\hline
			\textbf{Method} & \textbf{PSNR (dB)} & \textbf{SSIM} \\
			\hline
			Legitimate decoder (upper bound) & 28.4 & 0.882 \\
			Trained inversion network & 22.2 & 0.629 \\
			Optimization-based inversion & 6.8 & 0.043 \\
			\hline
		\end{tabular}
	\end{table}
	
	\section{Conclusion}
	\label{sec_6}
	This paper proposes a Swin Transformer-based semantic communication framework integrated with federated learning for privacy-preserving image transmission in UAV-assisted low-altitude networks. The hierarchical encoder-decoder architecture effectively extracts multi-scale semantic features under bandwidth-constrained UAV uplinks, consistently outperforming DeepJSCC and conventional separation-based baselines across AWGN, Rician, and Rayleigh fading channels. The federated aggregation mechanism enables collaborative model training without exposing raw imagery, yielding significant reconstruction gains over locally trained models with faster convergence and robust performance under non-IID data distributions. Privacy analysis shows that the framework provides strong resistance to gradient inversion during training and passive protection against semantic feature interception during inference, while remaining compatible with active defense mechanisms for scenarios with stringent privacy requirements.
	
	Future work will focus on extending the framework to higher-resolution UAV imagery, exploring adaptive aggregation strategies for large-scale UAV swarms, and integrating active inference-phase privacy defenses to provide end-to-end formal guarantees.
	
	\section*{Declaration of Competing Interest}
	
	The authors declare that they have no known competing financial interests or personal relationships that could have appeared to influence the work reported in this paper.
	
	\section*{Acknowledgements}
	
	This work was supported in part by the National Natural Science
	Foundation of China under Grants 62571450 and 62101450; the Key
	Research and Development Program of Shaanxi under Grant
	2025CY-YBXM-043; the Shanghai Academy of Spaceflight Technology
	under Grant SAST2025-037; the Doctoral Innovation Fund of
	Northwestern Polytechnical University under Grant CX2025078;
	the Open Fund of Intelligent Control Laboratory; the Open Fund
	of Key Laboratory of Radio Spectrum Testing Technology (The State
	Radio Monitoring Center Testing Center), Ministry of Industry and
	Information Technology; the National Science Foundation (NSF)
	under Grant ECCS-2302469; the Japan Science and Technology Agency
	(JST) Adopting Sustainable Partnerships for Innovative Research
	Ecosystem (ASPIRE) under Grant JPMJAP2326.
	
	\nocite{*}
	\bibliographystyle{cja}

\begin{thebibliography}{99}
		
		\bibitem{ref1}
		Lu K, Zhou Q, Li R, Zhao Z, Chen X, Wu J, et al. Rethinking modern communication from semantic coding to semantic communication. {\em IEEE Wirel Commun} 2023;30(1):158-64.
		
		\bibitem{ref10}
		G\"{u}nd\"{u}z D, Qin Z, Aguerri IE, Dhillon HS, Yang Z, Yener A, et al. Beyond transmitting bits: Context, semantics, and task-oriented communications. {\em IEEE J Sel Areas Commun} 2023;41(1):5-41.
		
		\bibitem{ref0}
		Yan Y, Zhang X, Li L, Lin W, Li R, Cheng W, et al. FSSC: Federated learning of transformer neural networks for semantic image communication. {\em GLOBECOM 2024 - 2024 IEEE Global Communications Conference}; 2024 Dec 8-12; Cape Town, South Africa. 2024. p. 1659-64.
		
		\bibitem{ref24}
		Xu S, Qi Y, Qi F, et al. FLSC-CI: Federated learning and semantic communication empowered multimodal terminal collaborative inferencing framework for IoT businesses. {\em IEEE Trans Netw Sci Eng} 2026. Forthcoming.
		
		\bibitem{ref25}
		Tong H, Yang Z, Wang S, Hu Y, Saad W, Yin C. Federated learning based audio semantic communication over wireless networks. {\em GLOBECOM 2021 - 2021 IEEE Global Communications Conference}; 2021 Dec 7-11; Madrid, Spain. 2021.
		
		\bibitem{ref2}
		Bourtsoulatze E, Burth Kurka D, G\"{u}nd\"{u}z D. Deep joint source-channel coding for wireless image transmission. {\em IEEE Trans Cogn Commun Netw} 2019;5(3):567-79.
		
		\bibitem{zeng2016energy}
		Zeng Y, Zhang R, Lim TJ. Wireless communications with unmanned aerial vehicles: Opportunities and challenges. {\em IEEE Commun Mag} 2016;54(5):36-42.
		
		\bibitem{LiChannel2026}
		Li M, Li L, Lin W, Han Z, Basar T. Beyond Gaussian assumptions: A general fractional HJB control framework for L\'{e}vy-driven heavy-tailed channels in 6G. {\em IEEE Trans Wirel Commun} 2026;25:7535-50.
		
		\bibitem{UAV3}
		Chen Y, Cheng W, Zhang W. Reconfigurable intelligent surface equipped UAV in emergency wireless communications: A new fading-shadowing model and performance analysis. {\em IEEE Trans Commun} 2024;72(3):1821-34.
		
		\bibitem{LinUAV2022}
		Lin W, Li L, Liu Y, He Y, Liu Y. Timeliness optimization of unmanned aerial vehicle lossy communications for Internet-of-Things. {\em Chin J Aeronaut} 2023;36(6):249-55.
		
		\bibitem{LiUAV2021}
		Li L, Sun Y, Cheng Q, Wang D, Lin W, Chen W. Optimal trajectory and downlink power control for multi-type UAV aerial base stations. {\em Chin J Aeronaut} 2021;34(9):11-23.
		
		\bibitem{UAV2}
		Thomas CK, Saad W. Neuro-symbolic causal reasoning meets signaling game for emergent semantic communications. {\em IEEE Trans Wirel Commun} 2024;23(5):4546-63.
		
		\bibitem{LiangUAV2024}
		Liang W, Wen S, Li L, Cui J, Fang F. Distributed user pairing and effective computation offloading in aerial edge networks. {\em Chin J Aeronaut} 2024;37(4):378-90.
		
		\bibitem{UAV4}
		Sarkar RR, Chakrabarty A, Rahman MZ. Low-end hand held communication devices in a post-disaster scenario. {\em 2022 14th International Conference on Computational Intelligence and Communication Networks (CICN)}; 2022 Dec 16-17; Al-Khobar, Saudi Arabia. 2022. p. 595-9.
		
		\bibitem{UAV5}
		Fu Y, Cheng W, Wang J, Yin L, Zhang W. Digital-analog transmission based emergency semantic communications. arXiv:2501.01616. 2025.
		
		\bibitem{ZhengLAN2026}
		Zheng Y, Li L, Lin W, Liang W, Du Q, Han Z. Optimal transport framework for ISAC in low-altitude networks: Joint resource allocation for cooperative communication and non-cooperative localization. {\em IEEE Trans Commun} 2026;74:1984-2000.
		
		\bibitem{ref11}
		Kountouris M, Pappas N. Semantics-empowered communication for networked intelligent systems. {\em IEEE Commun Mag} 2021;59(6):96-102.
		
		\bibitem{ref12}
		Yang W, Du H, Liew ZQ, Lim WYB, Xiong Z, Niyato D. Semantic communications for future Internet: Fundamentals, applications, and challenges. {\em IEEE Commun Surv Tutor} 2023;25(1):213-50.
		
		\bibitem{ref16}
		Yang M, Kim HS. Deep joint source-channel coding for wireless image transmission with adaptive rate control. {\em ICASSP 2022 - 2022 IEEE International Conference on Acoustics, Speech and Signal Processing}; 2022 May 23-27; Singapore. 2022. p. 5193-7.
		
		\bibitem{ref17}
		Xie H, Qin Z, Tao X, Letaief KB. Task-oriented multi-user semantic communications. {\em IEEE J Sel Areas Commun} 2022;40(9):2584-97.
		
		\bibitem{ref18}
		Kurka DB, G\"{u}nd\"{u}z D. DeepJSCC-f: Deep joint source-channel coding of images with feedback. {\em IEEE J Sel Areas Inf Theory} 2020;1(1):178-93.
		
		\bibitem{ref19}
		Yang K, Wang S, Dai J, Tan K, Niu K, Zhang P. WITT: A wireless image transmission transformer for semantic communications. {\em ICASSP 2023 - 2023 IEEE International Conference on Acoustics, Speech and Signal Processing}; 2023 Jun 4-10; Rhodes Island, Greece. 2023. p. 1-5.
		
		\bibitem{ref20}
		Yoo H, Jung T, Dai L, Kim S, Chae CB. Demo: Real-time semantic communications with a vision transformer. {\em ICC 2022 - IEEE International Conference on Communications Workshops}; 2022 May 16-20; Seoul, South Korea. 2022. p. 1-2.
		
		\bibitem{Zhang}
		Zhang K, Li L, Lin W, Yan Y, Li R, Cheng W, et al. Semantic successive refinement: A generative AI-aided semantic communication framework. {\em IEEE Trans Cogn Commun Netw} 2025;11(2):687-99.
		
		\bibitem{Yan}
		Yan Y, Li L, Zhang X, Lin W, Cheng W, Han Z. Adaptive semantic generation and NOMA-based interference-aware conveying for 6G networks. {\em IEEE Trans Wirel Commun} 2025;24(3):2404-16.
		
		\bibitem{ref22}
		Lim WYB, et al. Federated learning in mobile edge networks: A comprehensive survey. {\em IEEE Commun Surv Tutor} 2020;22(3):2031-63.
		
		\bibitem{YinFL2024}
		Yin T, Li L, Lin W, Ni T, Liu Y, Xu H, et al. Joint client scheduling and wireless resource allocation for heterogeneous federated edge learning with non-IID data. {\em IEEE Trans Veh Technol} 2024;73(4):5742-54.
		
		\bibitem{LiFL2022}
		Li L, Ma D, Ren H, Wang P, Lin W, Han Z. Toward energy-efficient multiple IRSs: Federated learning-based configuration optimization. {\em IEEE Trans Green Commun Netw} 2022;6(2):755-65.
		
		\bibitem{ref23}
		Shi G, Xiao Y, Li Y, Xie X. From semantic communication to semantic-aware networking: Model, architecture, and open problems. {\em IEEE Commun Mag} 2021;59(8):44-50.
		
		\bibitem{ref26}
		Vaswani A, Shazeer N, Parmar N, Uszkoreit J, Jones L, Gomez AN, et al. Attention is all you need. {\em Advances in Neural Information Processing Systems (NeurIPS)}; 2017 Dec 4-9; Long Beach, CA. 2017.
		
		\bibitem{ref27}
		Dosovitskiy A, Beyer L, Kolesnikov A, Weissenborn D, Zhai X, Unterthiner T, et al. An image is worth 16x16 words: Transformers for image recognition at scale. arXiv:2010.11929. 2020.
		
		\bibitem{ref28}
		Liu Z, et al. Swin Transformer: Hierarchical vision transformer using shifted windows. {\em IEEE/CVF International Conference on Computer Vision (ICCV)}; 2021 Oct 11-17; Montreal, QC, Canada. 2021.
		
		\bibitem{ref29}
		McMahan HB, Moore E, Ramage D, Hampson S, Arcas BAy. Communication-efficient learning of deep networks from decentralized data. {\em International Conference on Artificial Intelligence and Statistics (AISTATS)}; 2017 Apr 20-22; Fort Lauderdale, FL. 2017.
		
		\bibitem{mehta2021mobilevit}
		Mehta S, Rastegari M. MobileViT: Light-weight, general-purpose, and mobile-friendly vision transformer. {\em International Conference on Learning Representations (ICLR)}; 2022 Apr 25-29. 2022.
		
		\bibitem{ref_DLG}
		Zhu L, Liu Z, Han S. Deep leakage from gradients. {\em Advances in Neural Information Processing Systems (NeurIPS)}; 2019 Dec 8-14; Vancouver, Canada. 2019.
		
		\bibitem{ref_GradViT}
		Hatamizadeh A, Yin H, Roth H, Li W, Kautz J, Xu D, et al. GradViT: Gradient inversion of vision transformers. {\em IEEE/CVF Conference on Computer Vision and Pattern Recognition (CVPR)}; 2022 Jun 18-24; New Orleans, LA. 2022. p. 10011-20.
		
		\bibitem{ref_MIEA}
		Chen Y, Guo Z, Liang Y. The model inversion eavesdropping attack in semantic communication systems. {\em GLOBECOM 2023 - 2023 IEEE Global Communications Conference}; 2023 Dec 4-8; Kuala Lumpur, Malaysia. 2023. p. 1-6.
		
		\bibitem{ref_DP}
		Abadi M, Chu A, Goodfellow I, McMahan HB, Mironov I, Talwar K, et al. Deep learning with differential privacy. {\em Proceedings of the ACM SIGSAC Conference on Computer and Communications Security (CCS)}; 2016 Oct 24-28; Vienna, Austria. 2016. p. 308-18.
		
		\bibitem{ref_SecAgg}
		Bonawitz K, Ivanov V, Kreuter B, Marcedone A, McMahan HB, Patel S, et al. Practical secure aggregation for privacy-preserving machine learning. {\em Proceedings of the ACM SIGSAC Conference on Computer and Communications Security (CCS)}; 2017 Oct 30 - Nov 3; Dallas, TX. 2017. p. 1175-91.
		
		\bibitem{ref30}
		Krizhevsky A. Learning multiple layers of features from tiny images. Toronto: University of Toronto; 2009.
		
		\bibitem{ref31}
		Kingma DP, Ba J. Adam: A method for stochastic optimization. arXiv:1412.6980. 2014.
		
	\end{thebibliography}
	

\end{document}